\documentclass[]{spie}

\usepackage[]{graphicx}
\usepackage{caption}
\usepackage{subcaption}
\usepackage{xfrac}

\def  \be {\begin{equation}}
\def  \ee {\end{equation}}

\title{Focal Plane Alignment and Detector Characterization for the Subaru Prime Focus Spectrograph }

\author{Murdock  Hart\supit{a}, Robert H. Barkhouser\supit{a}, Michael Carr\supit{b}, Mirek Golebiowski\supit{a}, James E. Gunn\supit{b},  Stephen C. Hope\supit{a} and Stephen A. Smee\supit{a}
\skiplinehalf
\supit{a}Department of Physics and Astronomy, Johns Hopkins University, Baltimore, MD, 21218 (USA); \\
\supit{b}Department of Astrophysical Sciences, Princeton University, Princeton, NJ, 08544 (USA)
}

\begin{document} 
\maketitle

 %%%%%%%%%%%         ABSTRACT                   %%%%%%%%%%%%%%%%%%%

\begin{abstract}
 We describe the infrastructure being developed to align and characterize the detectors for the Subaru Measurement of Images and Redshifts (SuMIRe) Prime Focus Spectrograph (PFS).  PFS will employ four three-channel spectrographs with an operating wavelength range of 3800 $\AA$ to 12600 $\AA$. Each spectrograph will be comprised of two visible channels and one near infrared (NIR) channel, where each channel will use a separate Schmidt camera to image the captured spectra onto their respective detectors. In the visible channels, Hamamatsu 2k x 4k CCDs will be mounted in pairs to create a single 4k x 4k detector, while the NIR channel will use a single Teledyne 4k x 4k H4RG HgCdTe device. 
 
The fast f/1.1 optics of the Schmidt cameras will give a shallow depth of focus necessitating an optimization of the focal plane array flatness. The minimum departure from flatness of the focal plane array for the visible channels is set the by the CCD flatness, typically 10 $\mu$m peak-to-valley.  We will adjust the coplanarity for a pair of CCDs such that the flatness of the array is consistent with the flatness of the detectors themselves.   To achieve this we will use an optical non-contact measurement system to measure surface flatness and coplanarity at both ambient and operating temperatures, and use shims to adjust the coplanarity of the CCDs.
 
We will characterize the performance of the detectors for PFS consistent with the scientific goals for the project. To this end we will measure the gain, linearity, full well, quantum efficiency (QE), charge diffusion, charge transfer inefficiency (CTI), and noise properties of these devices. We also desire to better understand the non-linearity of the photon transfer curve for the CCDs, and the charge persistence/reciprocity problems of the HgCdTe devices. 
 
To enable the metrology and characterization of these detectors we are building two test cryostats nearly identical in design. The first test cryostat will primarily be used for the coplanarity measurements and sub-pixel illumination testing, and the second will be dedicated to performance characterization requiring flat field illumination. In this paper we will describe the design of the test cryostats. We will also describe the system we have built for measuring focal plane array flatness, and examine the precision and error with which it operates. Finally we will detail the methods by which we plan to characterize the performance of the detectors for PFS, and provide preliminary results.
\end{abstract}

 %%%%%%%%%%%         KEYWORDS                   %%%%%%%%%%%%%%%%%%%

\keywords{Optical Detectors, NIR Detectors, Detector Characterization, Focal Plane Array,  Subaru Telescope}

 %%%%%%%%%%%         INTRODUCTION                 %%%%%%%%%%%%%%%%%%%
\section{INTRODUCTION}
PFS is a multi-object spectrograph, that is four separate spectrographs, and it will be fiber-fed with 2400 fibers from Subaru's prime focus.  Each spectrograph will receive 600 fibers, and will consist of three channels covering the wavelength ranges 3800-6700 \AA, 6500-9700 \AA, and 9500-12600 \AA. Vacuum Schmidt cameras with a fast f/1.1 focal ratio will be used in each channel to focus the dispersed images of the fibers onto the detectors. The detectors for the two optical channels will be Hamamatsu\footnote{http://www.hamamatsu.com/us/en/index.html} 2K x 4K fully-depleted CCDs, with 15 $\mu$m pixels, operating at approximately 170 K. The prototyping of these devices was previously described in Kamata\cite{Kamata:2006hb} 2006 and their characterization for use in Hyper Suprime-Cam (HSC) was reported in Kamata\cite{Kamata:2010ic} 2010. A pair of the Hamamatsu detectors will be side-butted in each visible camera to yield a single 4K x 4K detector. The near infrared camera will use a 1.75 $\mu$m cutoff Teledyne\footnote{http://www.teledyne-si.com} H4RG-15 Mercury-Cadmium-Telluride (HgCdTe) device which is similarly a 4K x 4K device with 15 $\mu$m pixels. This detector will be operated at approximately 110 K. Blank\cite{Blank:2012db} provides a detailed report on the H4RG-15 prototypes functionality and performance test results.

 %%%%%%%%%%%          FPA Alignment                  %%%%%%%%%%%%%%%%%%%
\section{Focal Plane Alignment}
To use a 15 $\mu$m pixel size with a telescope as large as Subaru (8.2m) it will require a very fast beam, as the scale on the sky is determined by the cameras f/ratio. PFS will use a f/1.1 camera with fibers approximately 1.1 arcseconds in diameter, which will create a spot size on the detector of about 54 $\mu$m in diameter, or nearly 4 pixels in across. While the detectors will not be the dominate source of wavefront error in our instrument we still do desire to minimize any contribution which they may make to the total wavefront error of the instrument. The flatness of the Hamamatsu detectors first delivered for HSC was on the order of 20 $\mu$m peak-to-valley, but Hamamatsu has recently refined their process and been able to produce chips with flatness on the order of 10 $\mu$m peak-to-valley. The Teledyne detectors are specified to 20 $\mu$m peak-to-valley flatness, but their production chips have been found to typically be better than 12 $\mu$m peak-to-valley.

%% micro-epsilon confocal microscope
\subsection{Confocal Microscope}
\begin{figure}[h]
\centering
\begin{minipage}{.5\textwidth}
  \centering
  \includegraphics[width=\linewidth]{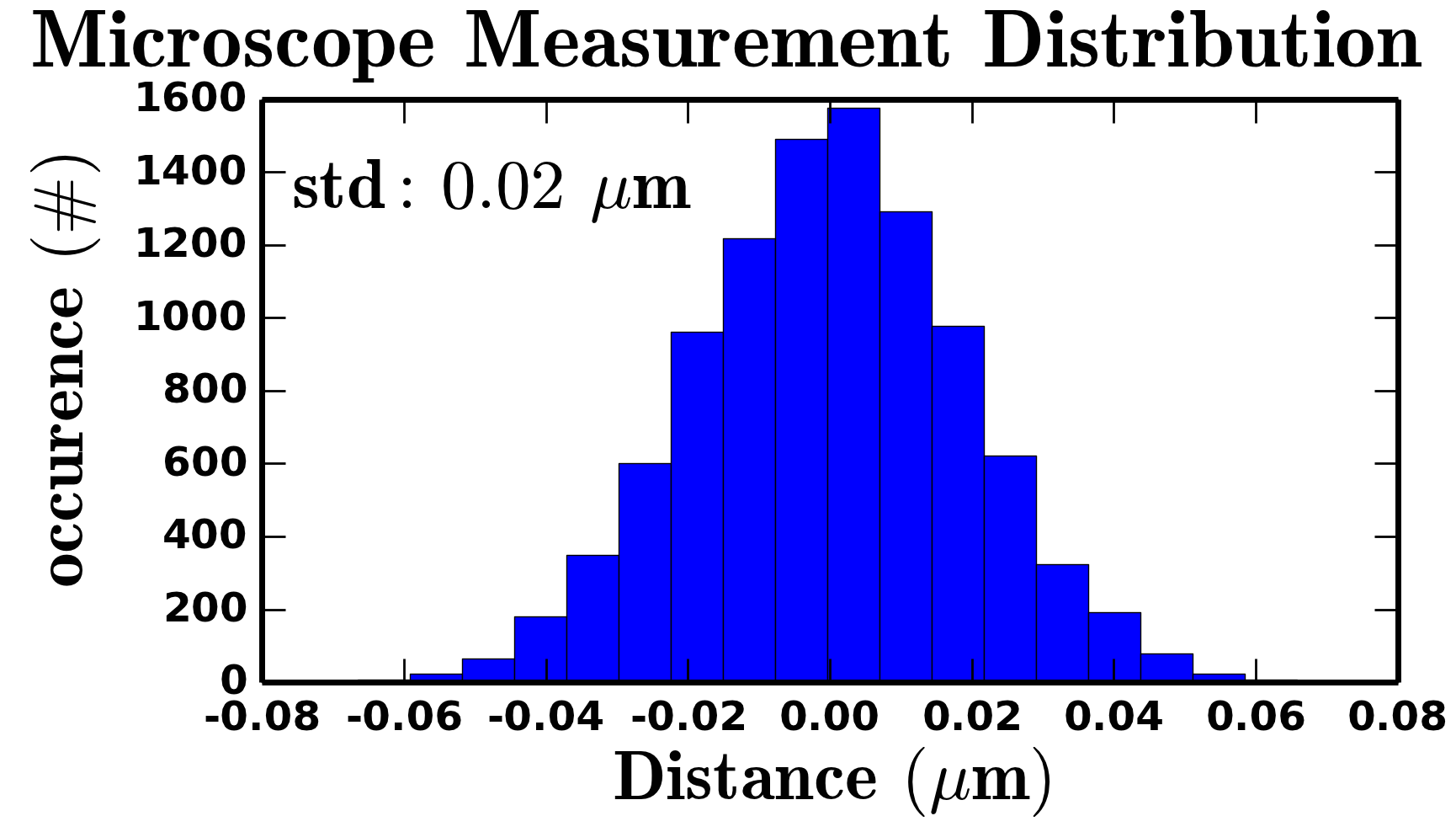}
  \captionsetup{width=0.9\linewidth}
  \captionof{figure}{Distribution of 10,000 confocal microscope measured distances with the microscope held at a fixed distance. The error on a microscope measured distance is on the order of 0.02 $\mu$m.}
  \label{micro_repeatability}
\end{minipage}%
\begin{minipage}{.5\textwidth}
  \centering
  \vspace{4.5mm}
  \includegraphics[width=0.97\linewidth]{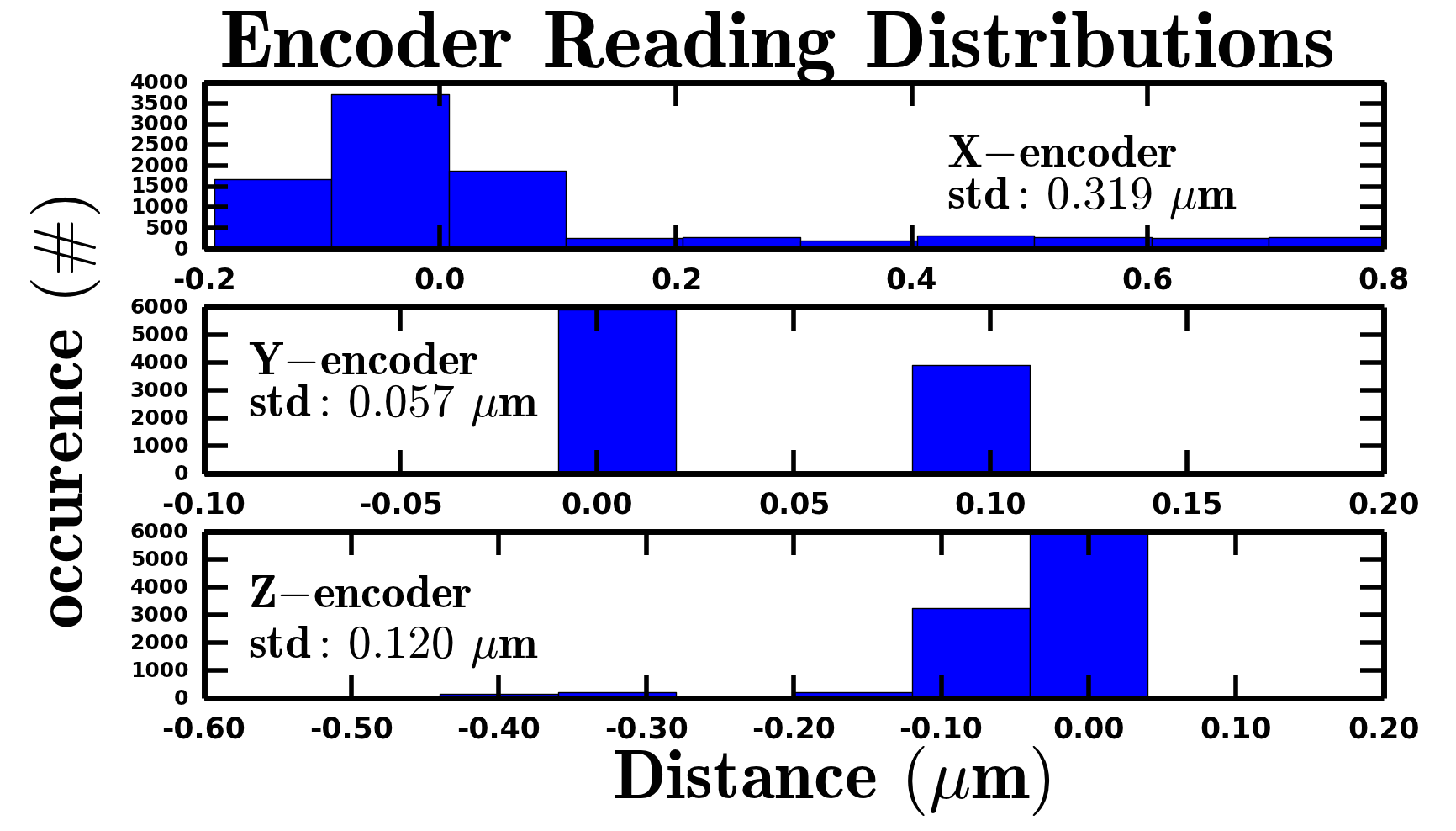}
  \captionsetup{width=0.9\linewidth}
  \captionof{figure}{Distribution of 10,000 CMM encoders measurements at a fixed position. The error on the CMM's X-axis was measured to be 0.32 $\mu$m, and is the dominant source of error for the non-contact measuring system.}
  \label{cmm_encoder_dist}
\end{minipage}
\end{figure}

The detector flatness and coplanarity needs to be measured non-contact and through a cryostat window. Derylo\cite{Derylo:2006kr} used a Micro-Epsilon\footnote{http://www.micro-epsilon.com/displacement-position-sensors/confocal-sensor/confocalDT\_2405/index.html} white light confocal microscope to measure detector topography for the Dark Energy Camera project. We decided to use a similar microscope made by Micro-Epsilon, a confocalDT IFS 2405 with a 10 mm effective measuring range, a 50 mm minimum measuring distance, and a 16 $\mu$m spot size. This device takes white light from a LED source and uses chromatic aberration to disperse light through a range of foci from the instrument head. The wavelength of the light focused at the surface of the target to be measured has the highest reflectance. The spectrum of reflected light is returned through the microscope head to be sensed by the microscope as a direct measure of the target surface's distance from the instrument head. The advertised accuracy of this device is approximately 2.5 $\mu$m  linearity over a 10 mm range with a maximum working distance of almost 60 mm, and a resolution of 60 nm. Fig. \ref{micro_repeatability} shows the distribution of measured values for 10,000 microscope reads with a fixed measuring distance. The error on the positional measurement for the microscope is on the order of 0.02 $\mu$m, which is small in comparison to the positioning error of the coordinate measuring machine (CMM).

%% positional measurement  error distribution plots

%% confocal microscope positioning
\subsection{Microscope Positioning}
To position the confocal microscope we are using a Wenzel\footnote{http://www.wenzelamerica.com} CMM with the microscope mounted parallel to the CMM's X-axis. The ability of this machine to follow in position is quite good, with an advertised precision on the order of $\pm3$ $\mu$m. Fig.\ref{cmm_encoder_dist} is the distribution of 10,000 encoder readings at a fixed position, which shows that the encoder error on all three axes are sub-micron. The root square sum of distance measurement errors of the non-contact measuring system is dominated by the CMM's X-axis error of 0.32 $\mu$m, and is the primary uncertainty in the non-contact distance measurements.

\begin{figure}[h]
\centering
\begin{minipage}{.5\textwidth}
  \centering
  \includegraphics[width=\linewidth]{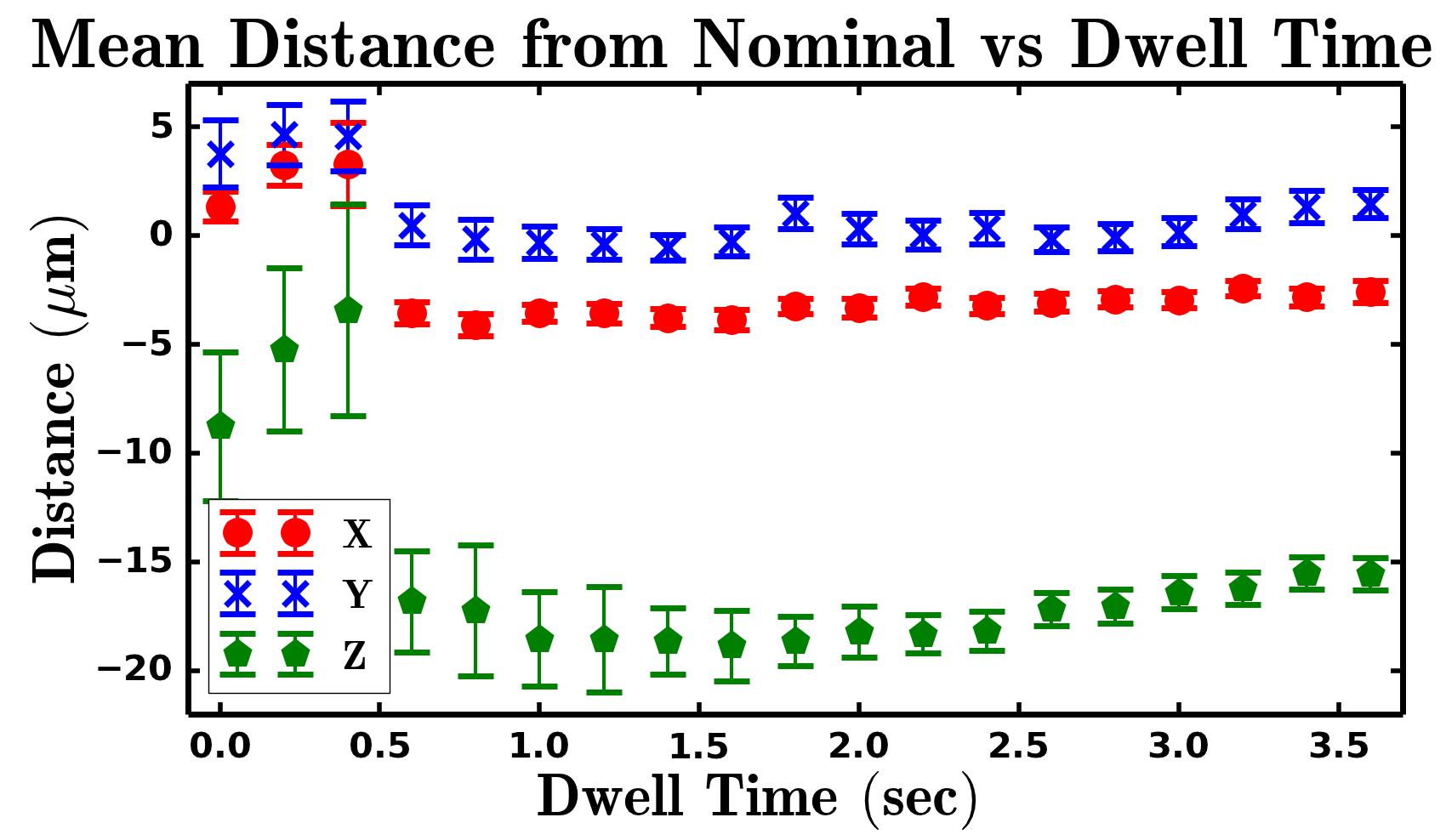}
  \captionsetup{width=0.9\linewidth}
  \captionof{figure}{CMM absolute position and error as a function of delay before reading the encoders. }
  \label{mean_dist_vs_dwell}
\end{minipage}%
\begin{minipage}{.5\textwidth}
  \centering
  %\vspace{2mm}
  \includegraphics[width=\linewidth]{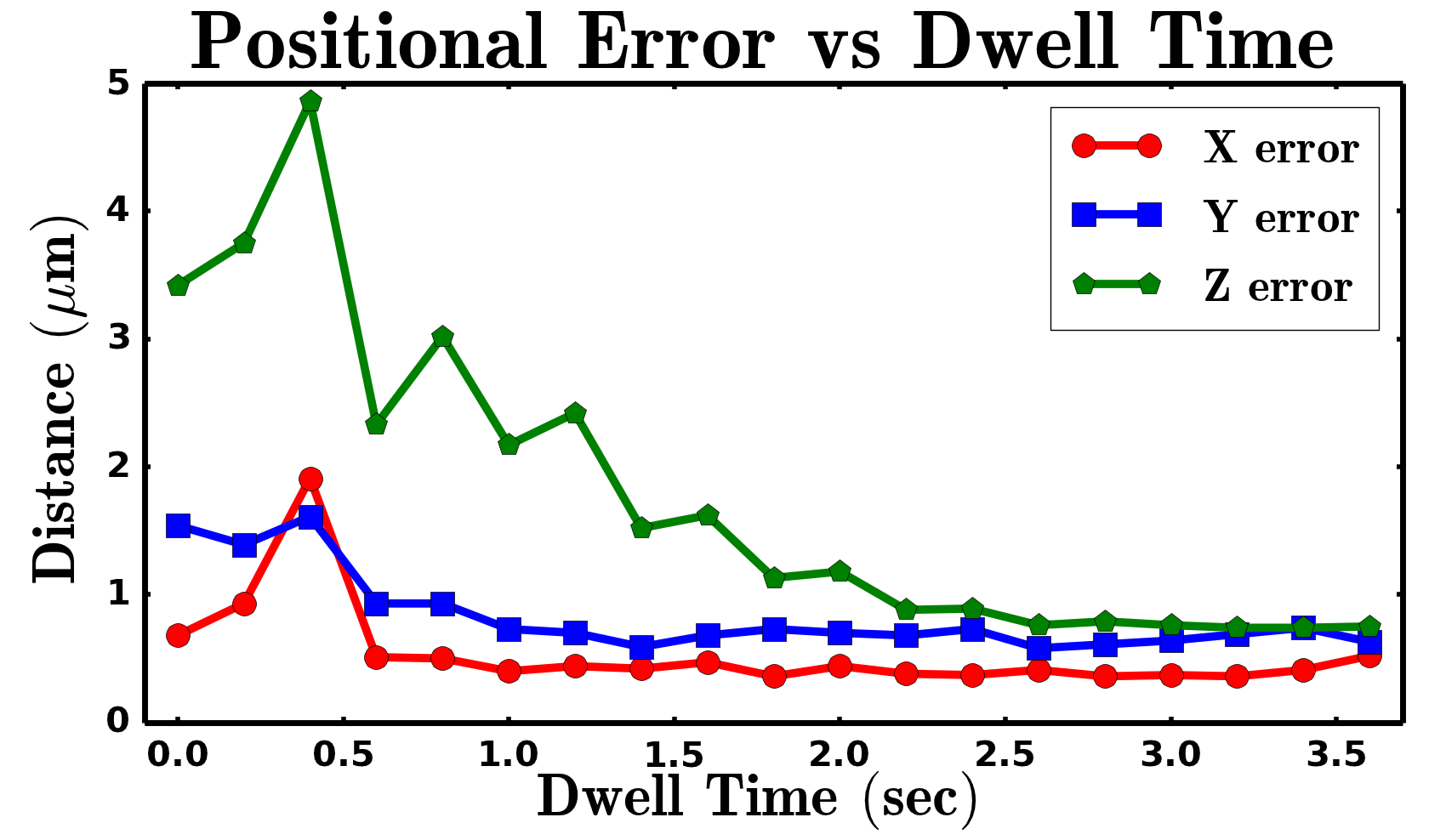}
  \captionsetup{width=0.9\linewidth}
  \captionof{figure}{CMM absolute positional error as a function of delay before reading the encoders.}
  \label{positional_error_vs_dwell}
\end{minipage}
\end{figure}

The absolute positioning of the CMM is much less precise. To control the absolute position of an individual axis a PID feedback loop is employed with no derivative term. Fig.\ref{mean_dist_vs_dwell} is the absolute position, nominal minus actual position, for 4000 movements versus a delay, or dwell time before reading the encoders. Fig.\ref{mean_dist_vs_dwell} shows that the CMM is continuously correcting for absolute position. Also Fig.\ref{mean_dist_vs_dwell} shows the X and Y axes having an overshoot, as would be expected with a PID feedback loop, but a large negative offset in absolute positioning is seen in the Z-axis. It has been surmised that the negative offset in the Z-axis is a consequence of this axis having a force applied to it from not only the servo motor but also from gravity. Fig.\ref{positional_error_vs_dwell} is the absolute positional error for 4000 movements versus dwell time. The X-axis, the axis upon which the confocal microscope is mounted, shows to have minimal error immediately after arriving at a position, and all further non-contact surface measurements were conducted using zero dwell time.

%% absolute positioning  error distribution plots

%% detector surface topography
\subsection{Detector Topography}
\begin{figure}[h]
	\hspace{-1.4cm}
	\includegraphics[width=0.95\textwidth]{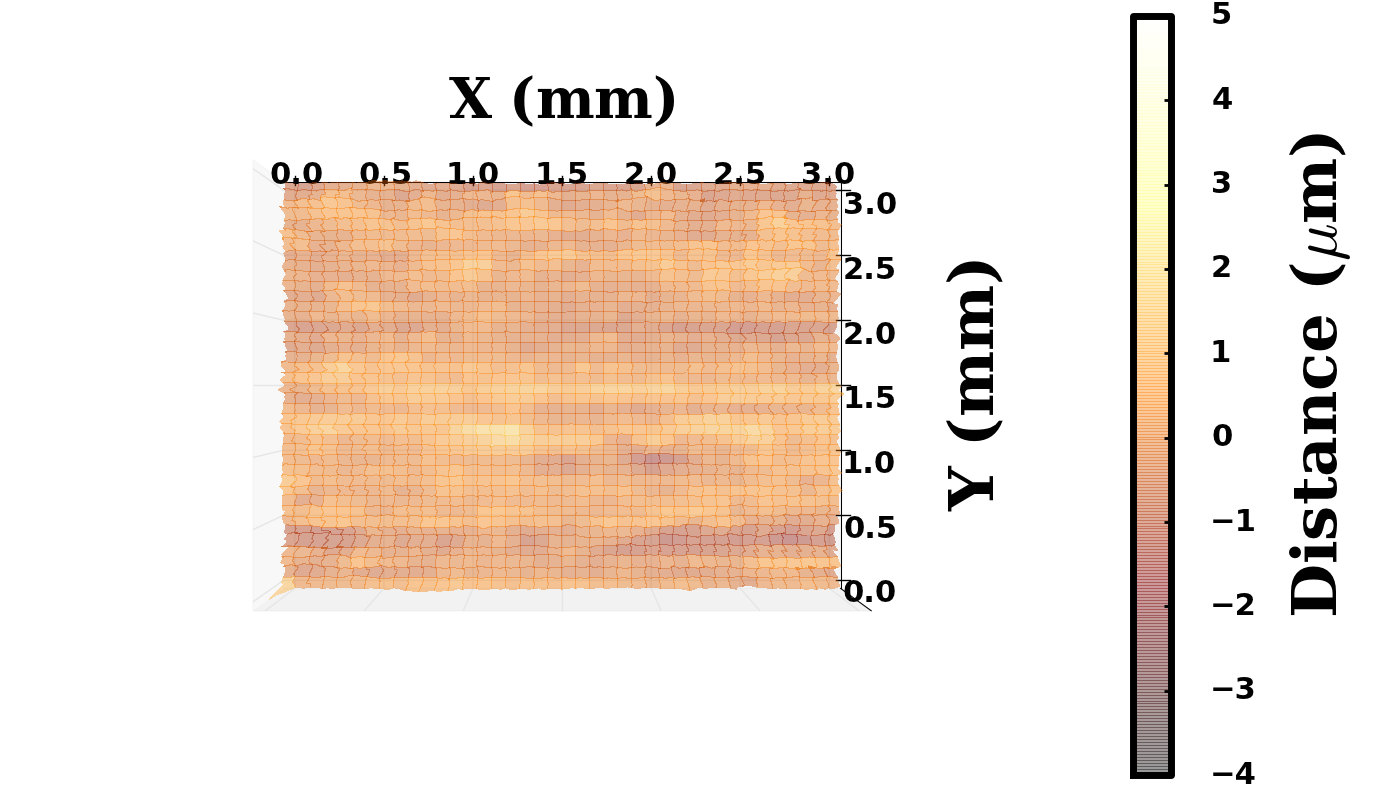}
	\vspace{2mm}
	\caption{Topographic plot taken over a randomly selected 3 mm square sub-sample of a Hamamatsu CCD with a 25 $\mu$m measuring pitch. The peak-to-valley flatness is on the order of 4 $\mu$m, and the RMS surface flatness was found to be approximately 0.5 $\mu$m. }
	\label{topography_dense}
\end{figure}

\begin{figure}[h]
\begin{minipage}{.5\textwidth}
 \centering
  %\vspace{3mm}
  \includegraphics[width=\linewidth]{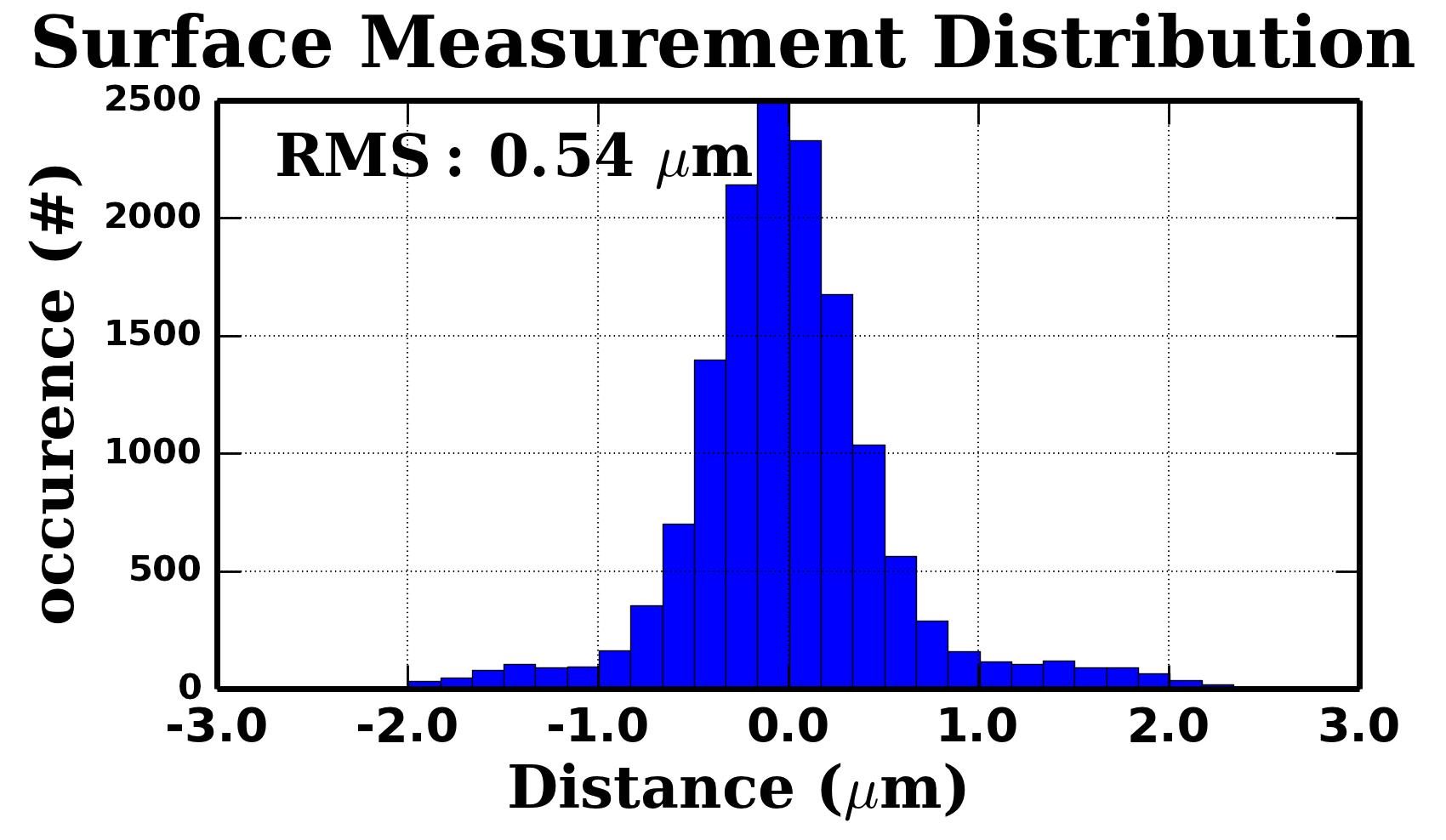}
  \captionsetup{width=0.9\linewidth}
  \captionof{figure}{Distribution of surface topographic measurements taken over a randomly selected 3 mm square sub-sample of the CCD with a 25 $\mu$m measurement sampling pitch. The peak-to-valley deviation from flatness is on the order 4 $\mu$m, while the surface RMS was approximately 0.5 $\mu$m. }
  \label{ccd_dense_sampling_dist}
\end{minipage}%
\begin{minipage}{.5\textwidth}
\centering
\vspace{-8mm}
 \includegraphics[width=\linewidth]{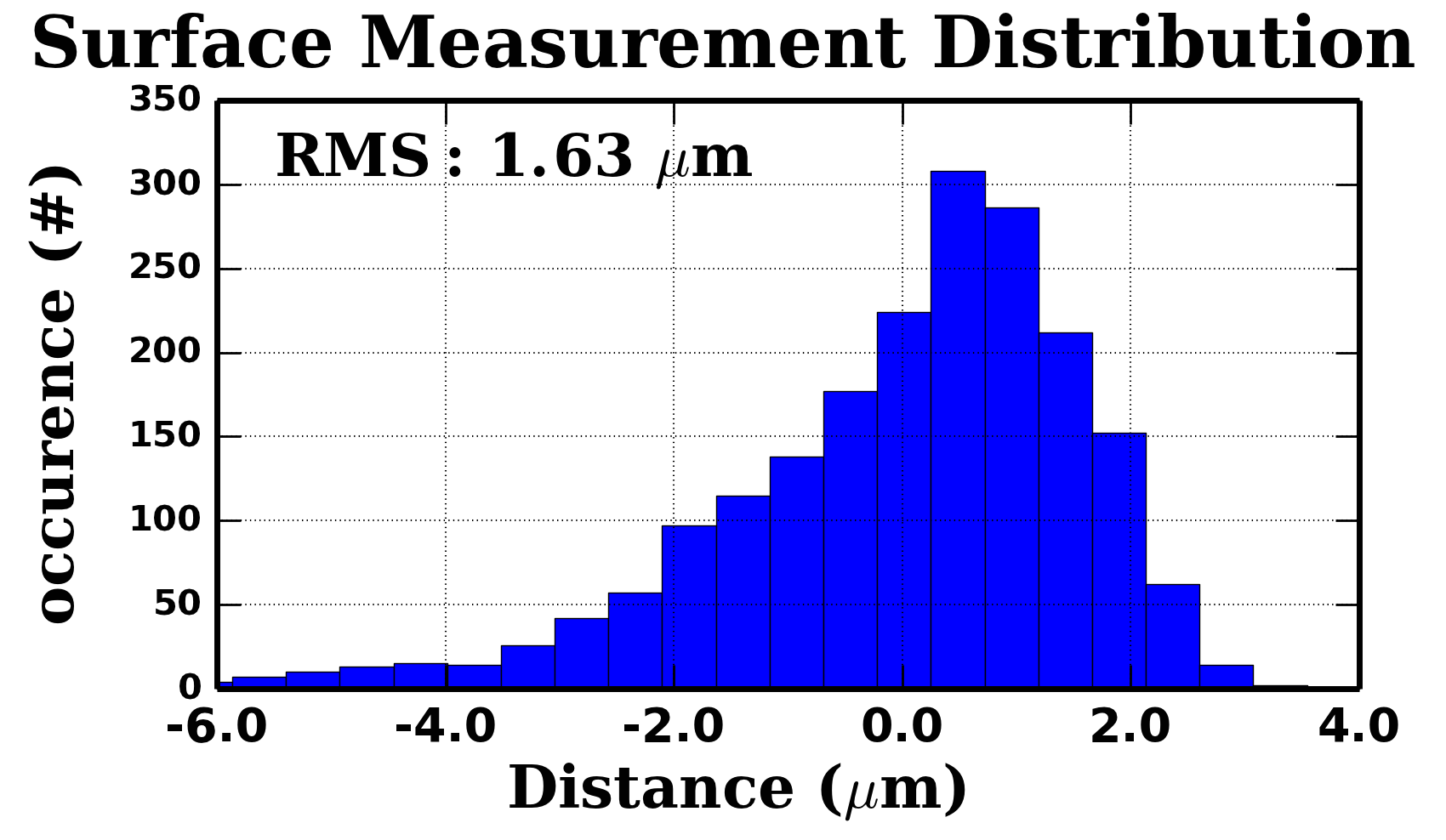}
  \captionof{figure}{Distribution of surface topographic measurements taken over the full extent of the CCD with a 1 mm measurement sampling pitch. The peak-to-valley deviation from flatness is on the order 10 $\mu$m. }
 \label{ccd_full_surface_dist}
\end{minipage}
\end{figure}

A 2D array of  points taken over a randomly selected 3 mm square sub-sample of the CCD using a 25 $\mu$m sampling pitch was used to examine the short scale structure of the surface. Fig.\ref{topography_dense} is a topographic plot taken to sample the short scale structure, using these measurements an analysis of the distribution of measured distances was performed to look for outliers. Fig.\ref{ccd_dense_sampling_dist} is the distribution of surface topographic measurements taken with the 25 $\mu$m measuring pitch, and shows on short length scales the peak-to-valley surface flatness is on the order of 4 $\mu$m, and the RMS surface flatness is sub-micron.

\begin{figure}[h]
	\hspace{-1.4cm}
	\includegraphics[width=0.95\textwidth]{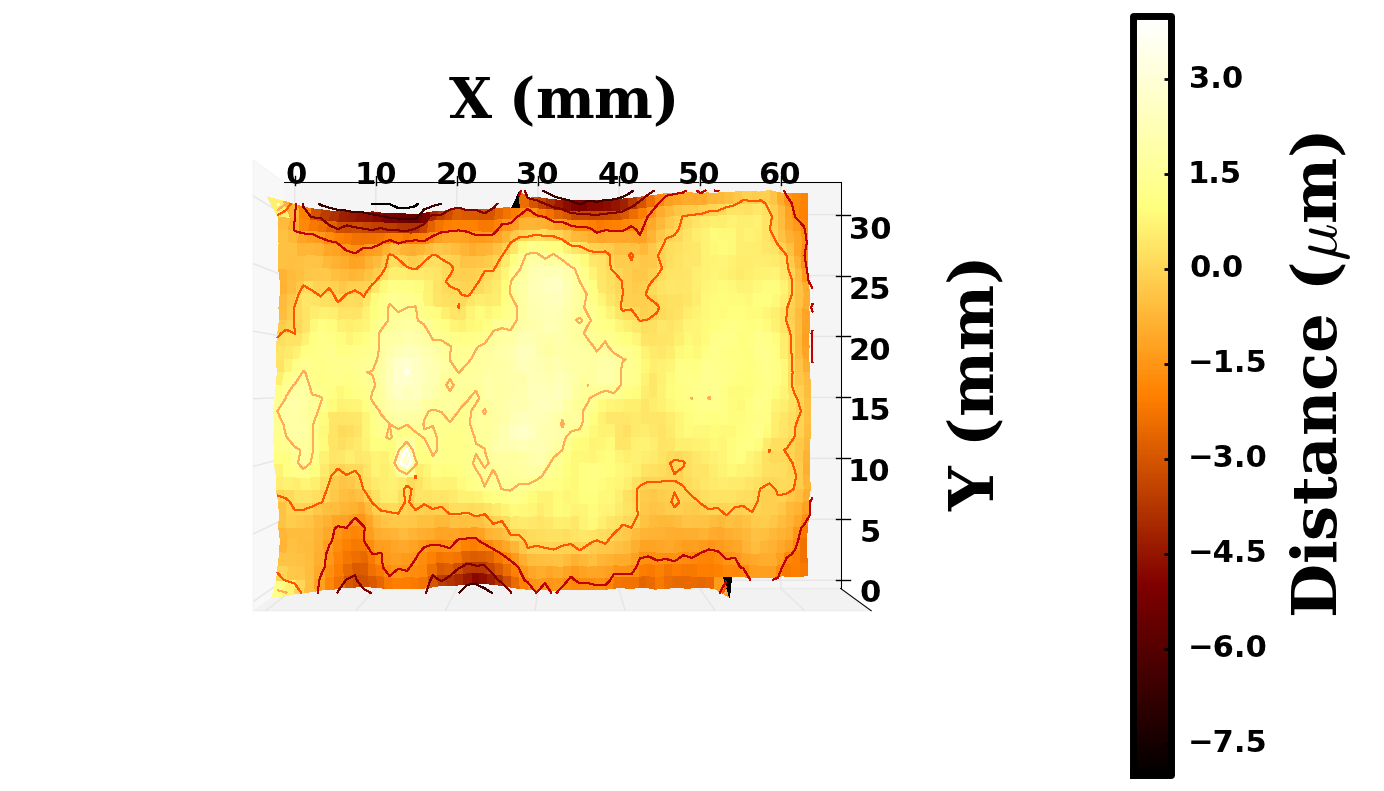}
	\vspace{2mm}
	\caption{Topographic plot taken over the entire extant of the CCD with a 1 mm measurement sampling pitch. The peak-to-valley flatness is on the order of 10 $\mu$m, while the surface RMS was approximately 1.5 $\mu$m.}
	\label{ccd_surface_plot}

\end{figure}

Fig.\ref{ccd_surface_plot} is a contour plot of an HSC mechanical Hamamatsu CCD taken of the full extent of the detector at a 1 mm sampling pitch. While Fig.\ref{ccd_full_surface_dist} is the distribution of topographic surface measurements from Fig.\ref{ccd_surface_plot}. The CCDs measured thus far have peak-to-valley flatness on the order of 10 $\mu$m, and RMS surface flatness on the order of 2 $\mu$m. An examination of Fig.\ref{ccd_surface_plot} shows the largest deviations in flatness are near the edges of the chip, and Fig.\ref{ccd_full_surface_dist} shows that the bulk of measured points are between $\pm$2 $\mu$m. Further measurements of the surface topography will be performed at operating temperatures.

%% topography dense sampling

%% surface topography distribution

%% surface topography contour plots

%% coplanar adjustment
\subsection{Coplanar Adjustment}
\begin{figure}[h!]
	\centering
	\includegraphics[width=0.8\textwidth]{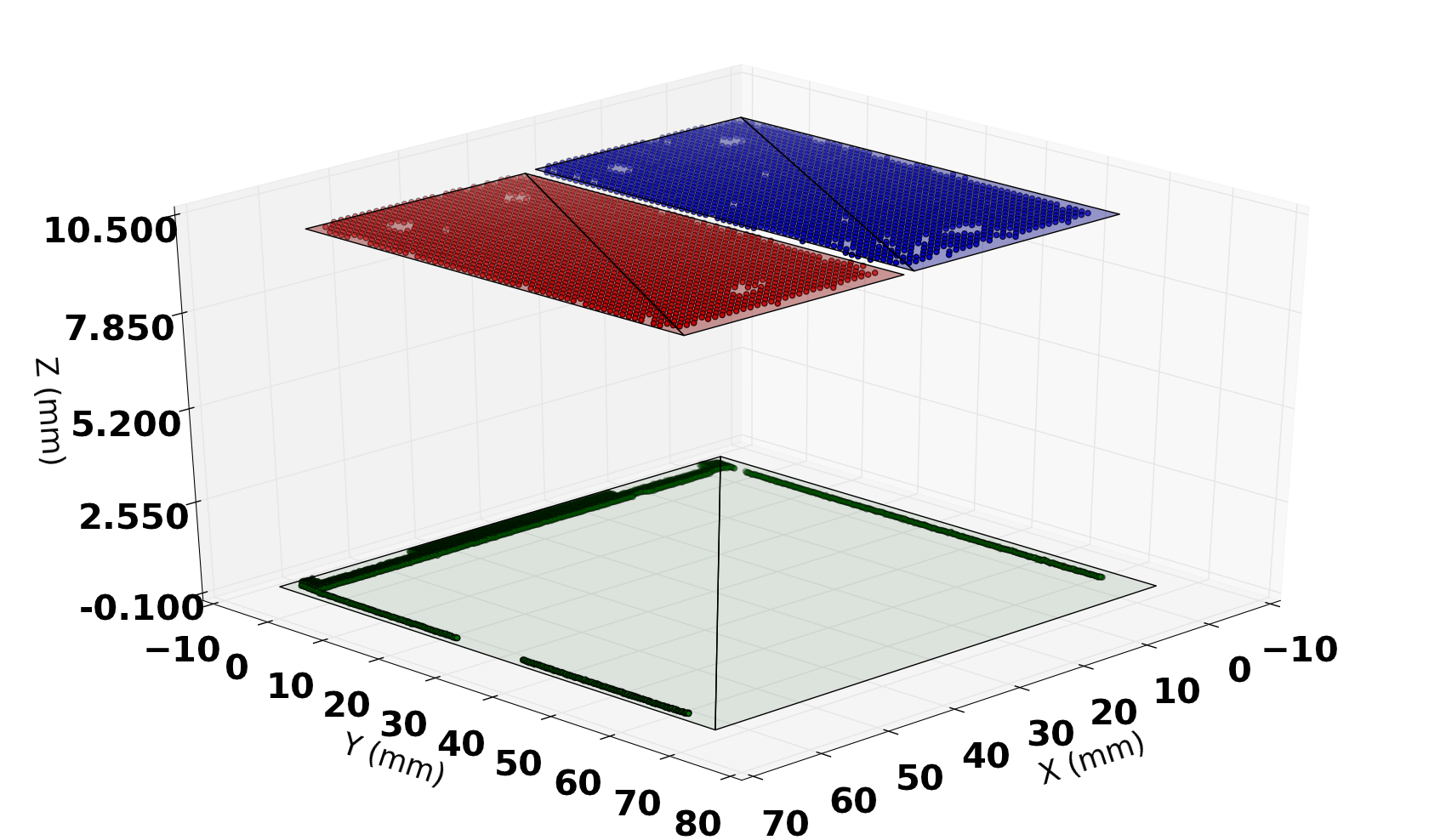}
	\vspace{4mm}
	\caption{Coplanar measurement for a pair of Hamamatsu CCDs with reference measurement taken from the mounting plate. The mounting plate is fitted by the green plane at the bottom, and the CCDs are individually fit with the blue and red planes above the backplate. For PFS the active surface of the detectors is specified to be 10 mm away from the mounting plate. }
	\label{coplanar_w_backplate}
\end{figure}

For coplanarity measurements a 2D array of points using a 1 mm step size will be sampled across the entire extent of the detectors. A median plane will be fit through the measured points of each Hamamatsu detector, and it is these reference planes that will be aligned to be coplanar by shimming the detectors on their mounting plate.  Fig.\ref{coplanar_w_backplate} shows the measurement of a pair of CCDs with respect to their mounting plate. The active surfaces of the detectors are specified to be 10 mm away from the mounting surface.

%% raw coplanar measurement show both CCDs and the backplate	

\begin{figure}[h]
\centering
\begin{minipage}{.5\textwidth}

  \hspace{-1.5cm}
  \includegraphics[width=1.4\linewidth]{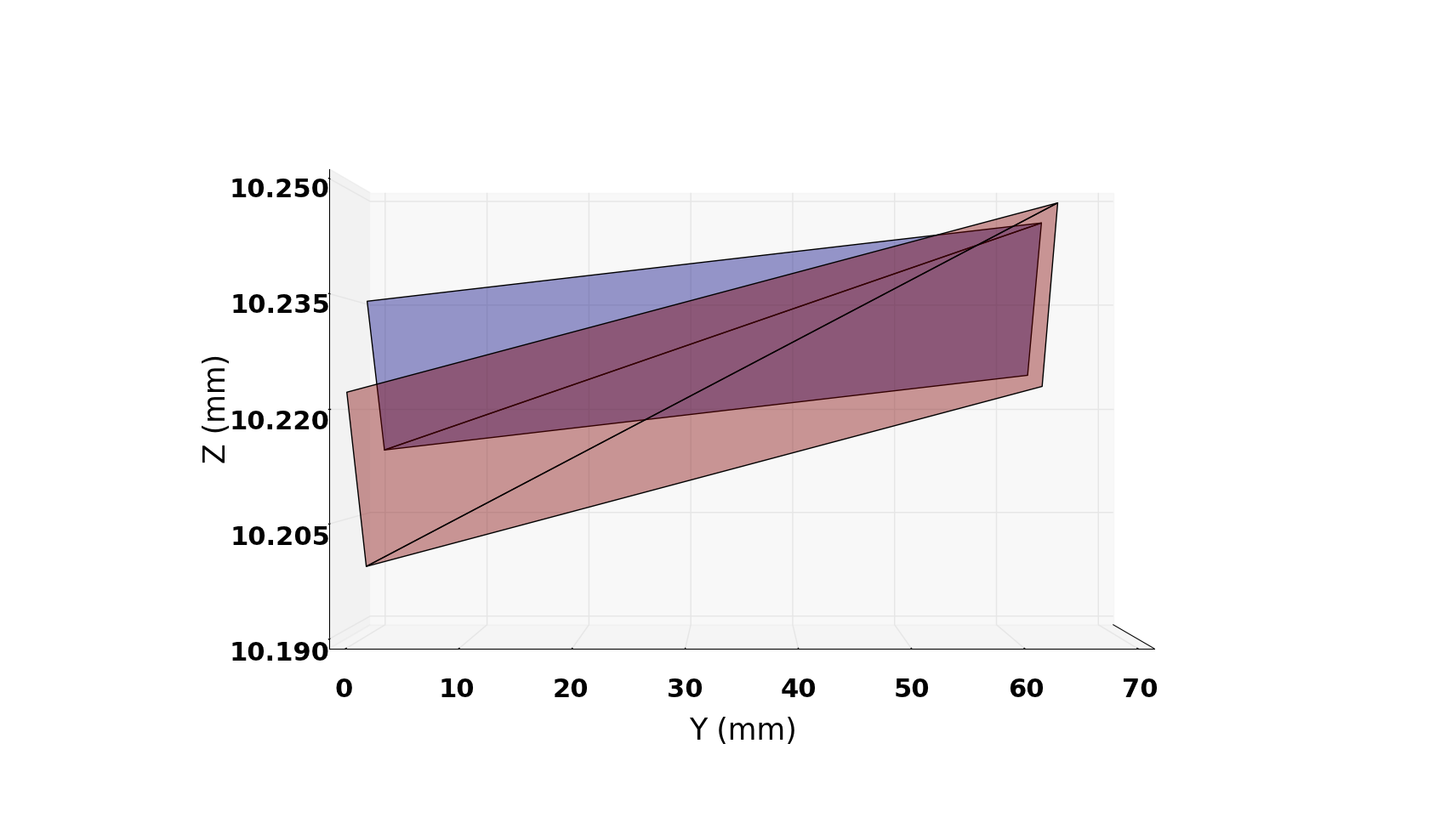}
  \captionsetup{width=0.9\linewidth}
  \captionof{figure}{Side view of CCDs with nominal 900 $\mu$m thick shims. }
  \label{before_side_view}
\end{minipage}%
\begin{minipage}{.5\textwidth}
  \hspace{-1.2cm}
  \vspace{-1mm}
  \includegraphics[width=1.4\linewidth]{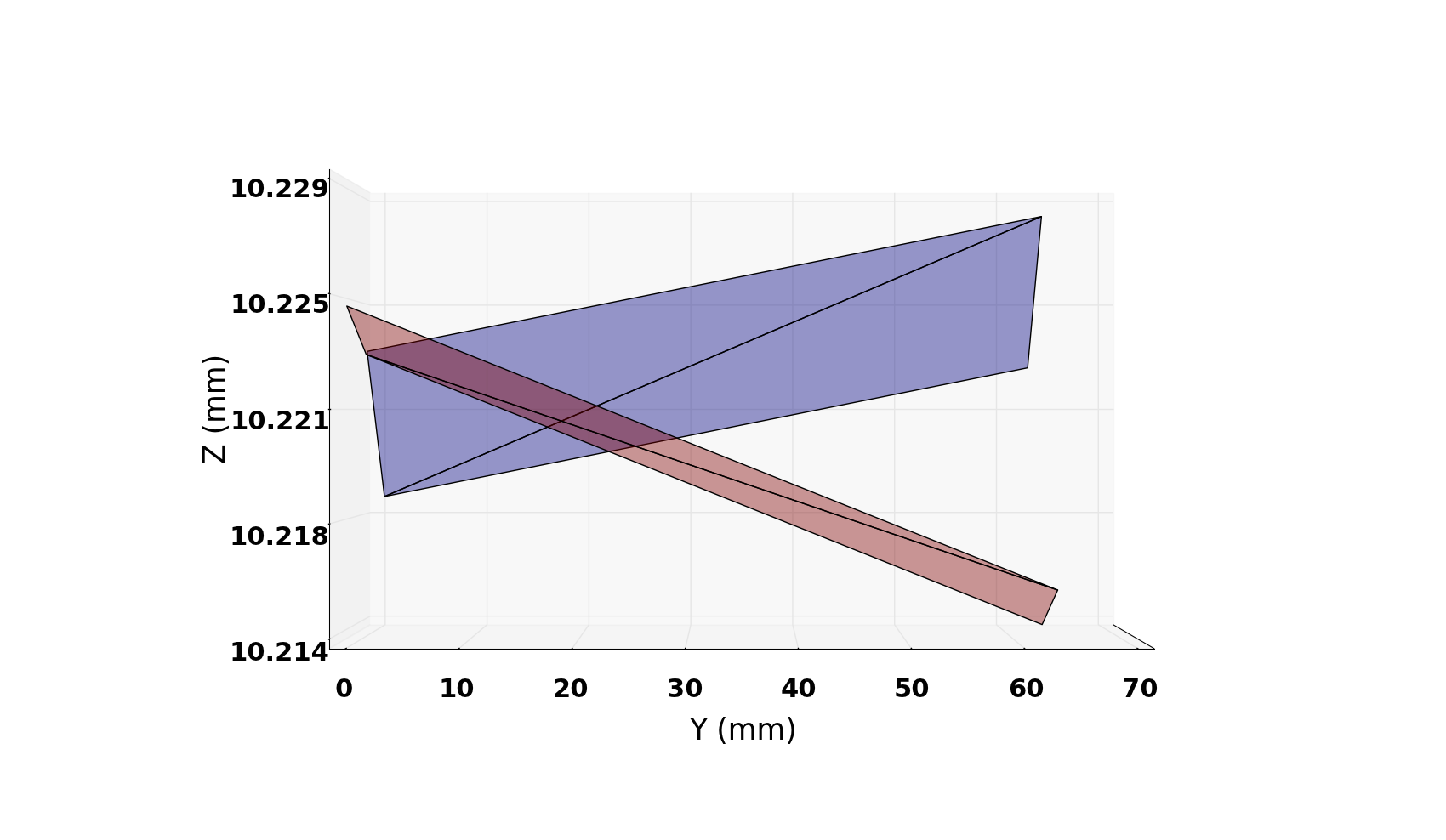}
  \captionsetup{width=0.9\linewidth}
  \captionof{figure}{Side view of CCDs after one shimming correction.}
  \label{after_side_view}
\end{minipage}
\end{figure}

\begin{figure}[h]
\centering
\begin{minipage}{.5\textwidth}
  \hspace{-1.8cm}
  \includegraphics[width=1.4\linewidth]{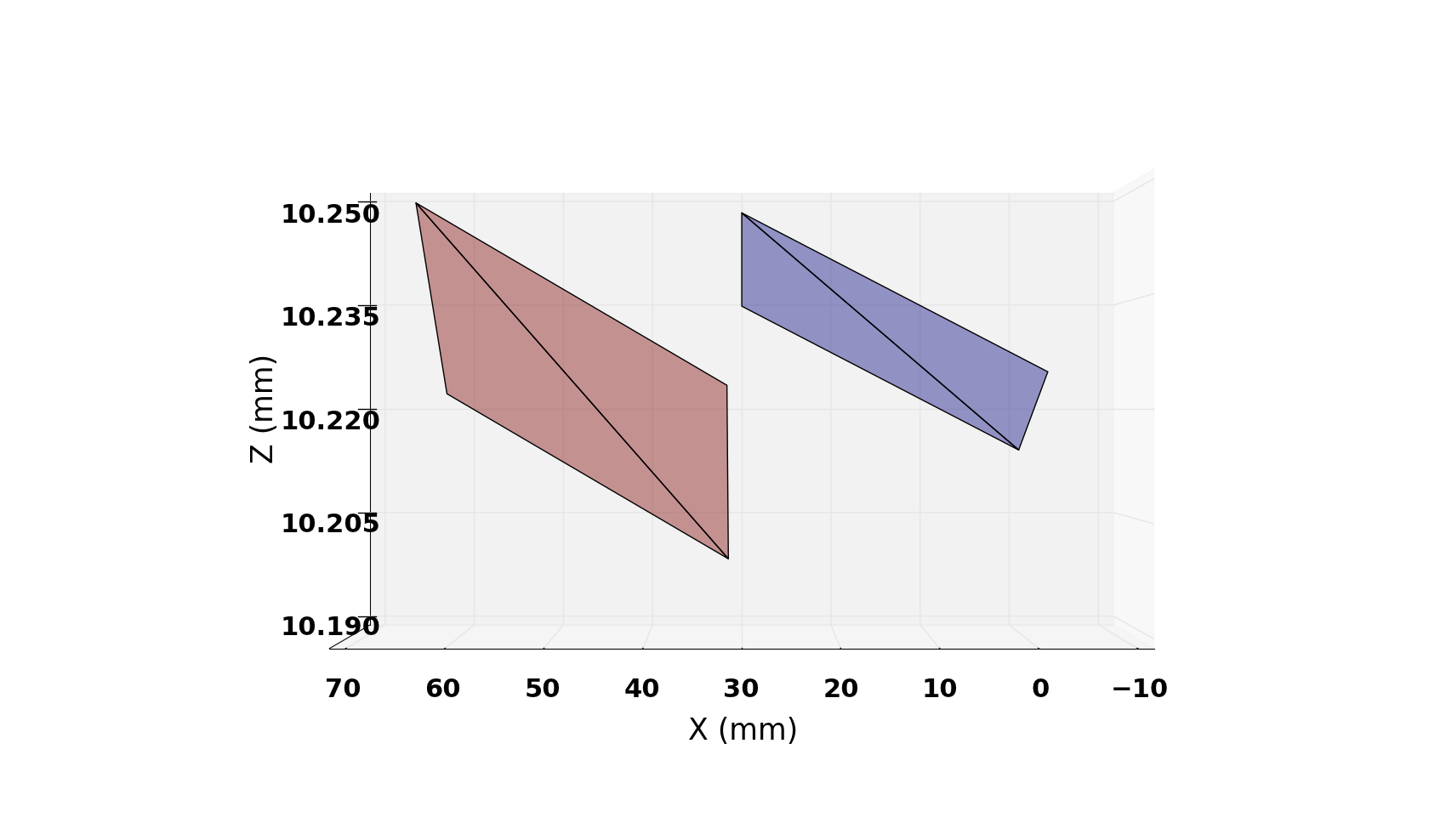}
  \captionsetup{width=0.9\linewidth}
  \captionof{figure}{End view of CCDs with nominal 900 $\mu$m thick shims. }
  \label{before_end_view}
\end{minipage}%
\begin{minipage}{.5\textwidth}
  \hspace{-1.5cm}
  %\vspace{-2mm}
  \includegraphics[width=1.4\linewidth]{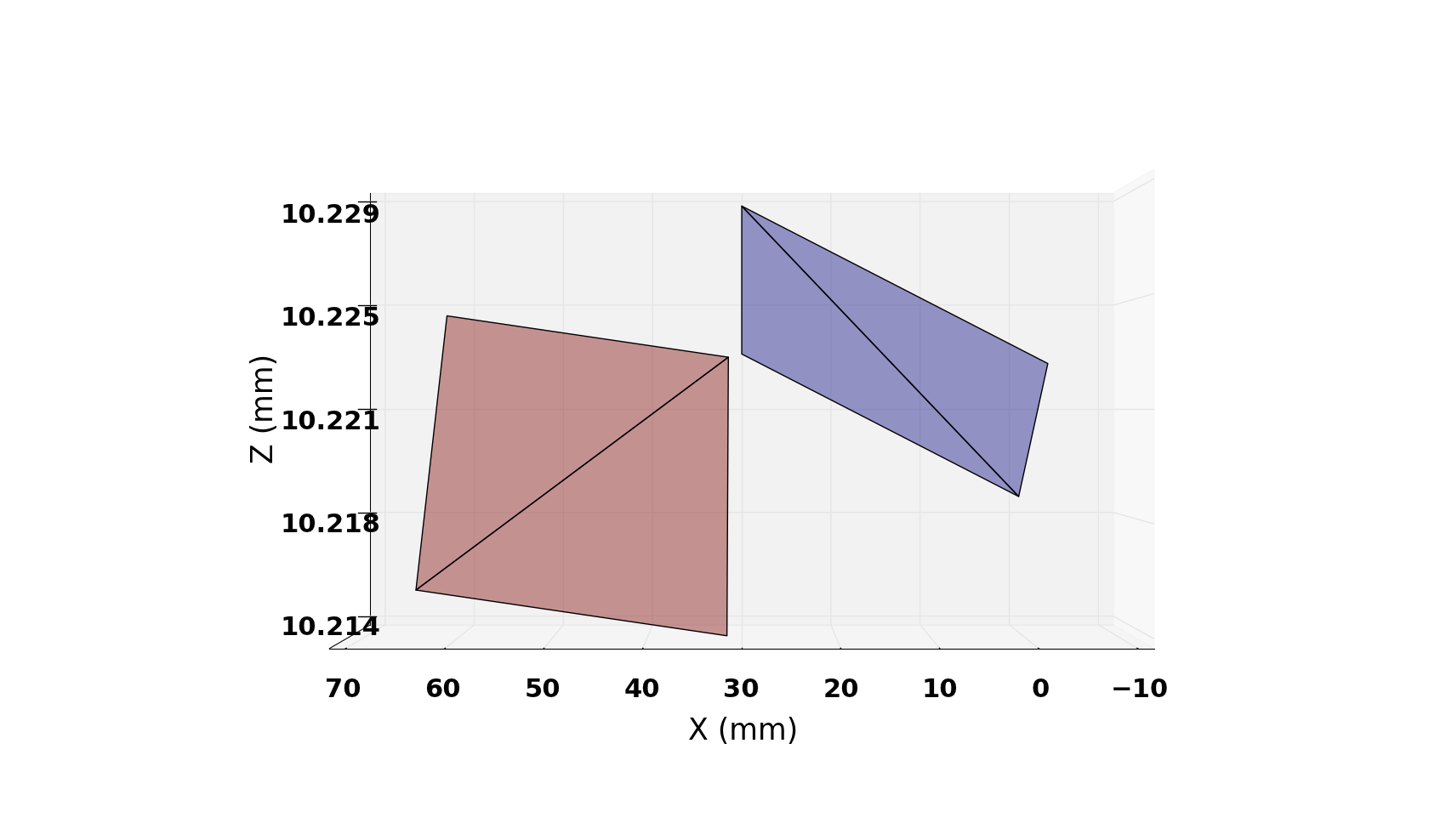}
  \captionsetup{width=0.9\linewidth}
  \captionof{figure}{End view of CCDs after one shimming correction.}
  \label{after_end_view}
\end{minipage}
\end{figure}

For PFS the specification for the coplanarity of the detectors shall be such that the RMS of the detector pair shall not exceed 4 $\mu$m with respect to their best fit planes. Initially shims with a range of sizes stepping in 10 $\mu$m increments were fabricated, but upon measuring the shims very little scatter was found in each shim size. New sizes of shims were fabricated, but now having a 5 $\mu$m step size giving the resolution necessary to meet PFS's coplanarity requirement. Fig.\ref{before_side_view} is the side view of the as measured detectors with nominal 900 $\mu$m thick shims installed, and Fig.\ref{after_side_view} is the side view after one shimming correction has been performed. While Fig.\ref{before_end_view} and Fig.\ref{after_end_view} are the end views of the before and after one shim correction has been performed respectively.

%% CCD side view before and after shimming 

%% CCD end view before and after shimming 

 %%%%%%%%%%%         TEST Cryostat                 %%%%%%%%%%%%%%%%%%%
\section{Test Cryostat}
\begin{figure}[h]
	\centering
	\includegraphics[width=0.6\textwidth]{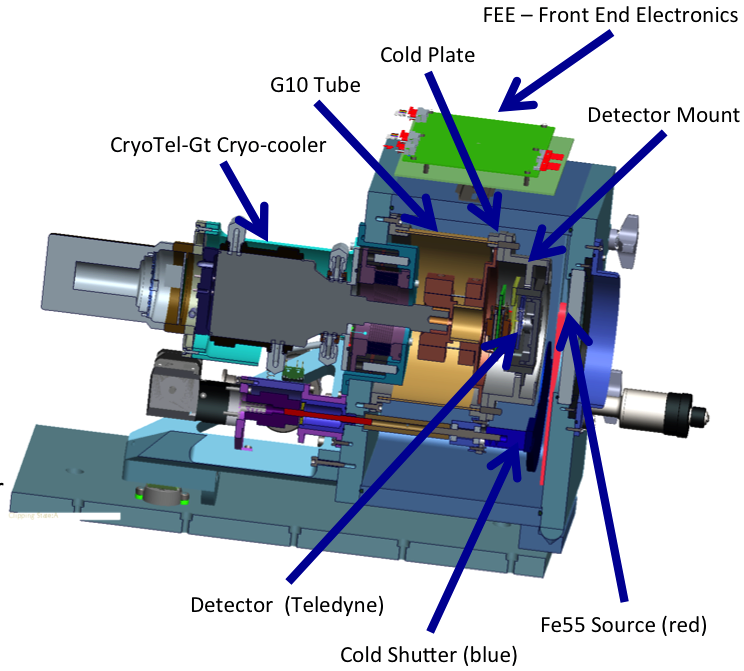}
	\caption{Cross section view of the test cryostat. }
	\label{tcryo_xsec}
\end{figure}

The initial design criteria for the test cryostat were: 1- needed to have the ability to fully flat field illuminate detectors on the order of 65 mm$^{2}$ in size which requires a window of greater than 100 mm clear aperture. 2- needed a cold shutter to facilitate dark exposures for the NIR detector. 3- have a minimal distance from the outside surface of the window to the detector active surface to allow for coplanarity measurements and sub-pixel illumination, with less than 30 mm being targeted. 4- needed to be able evaluate the cooling and lifetime performance of the Sunpower CryoTel-GT\footnote{http://www.sunpowerinc.com/cryocoolers/gt.php} cryo-coolers, and to further test an active damping system currently being developed. 5- the test cryostat needed to be portable  to allow for movement between configurations. With these criteria in mind a short search was performed for an off the shelf unit, and it was quickly realized no such item existed. With these initial criteria we began to design a test cryostat to meet these specifications, and a cross section view of the resulting design is shown in Fig. \ref{tcryo_xsec}. At a later time it was decided to add a mechanism by which to allow for x-ray illumination using a Fe55 source, and that the test cryostat would also be used to test the prototype camera electronics. The requirement for portability was dropped in favor of building two cryostats, and leaving them setup in their respective configurations.

%% cross section view of test cryostat	

\begin{figure}[h]
\centering
\begin{minipage}{.47\textwidth}
  \centering
  \includegraphics[width=0.9\linewidth]{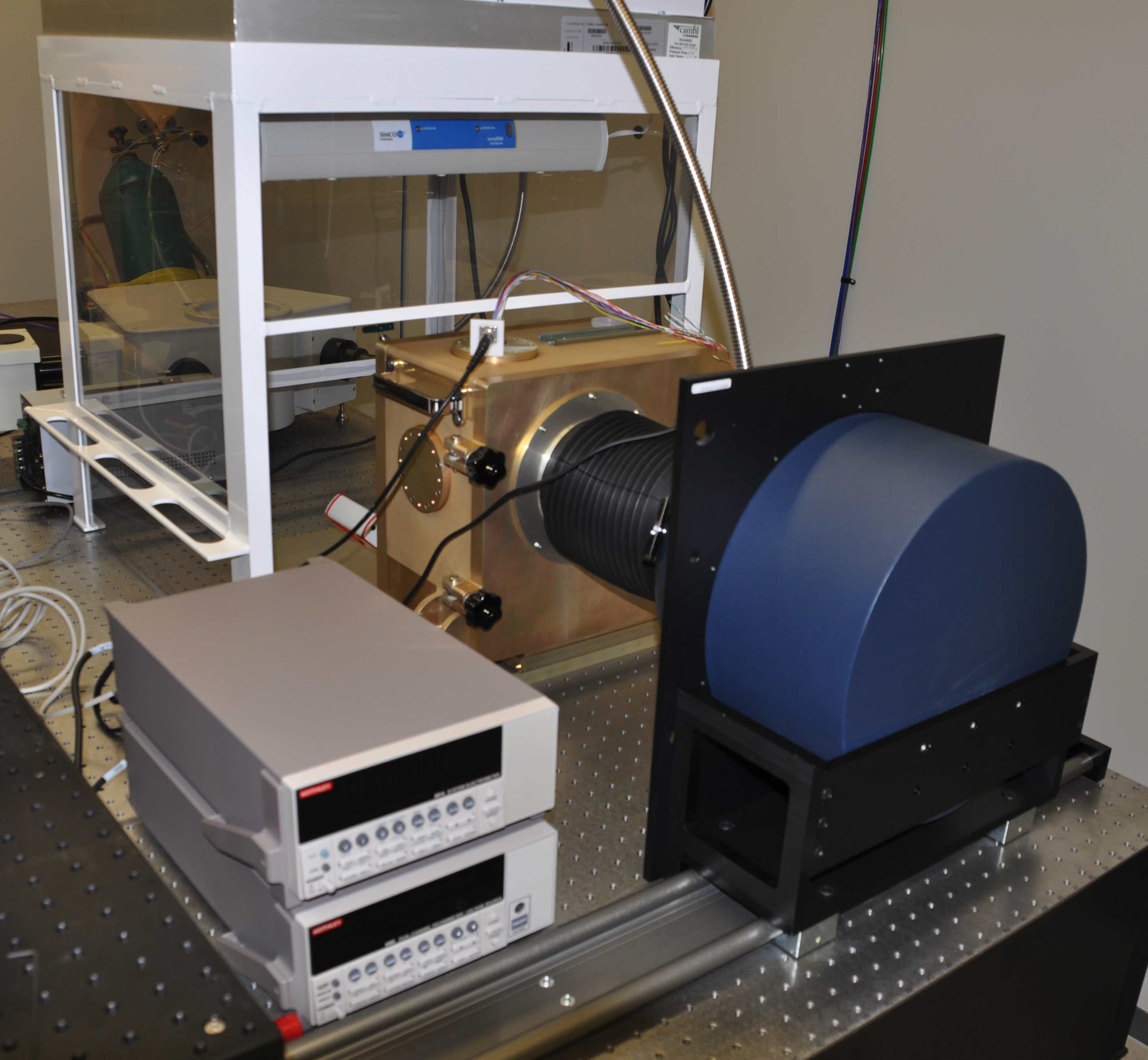}
  \vspace{2mm}
  \captionsetup{width=0.9\linewidth}
  \captionof{figure}{Photo showing the flat fielding setup, with the test cryostat, and Labsphere integrating sphere, mounted on a optical bench. Inside the dewar is an OSI Optoelectronics photodiode being used to calibrate the flux at the detector versus at the calibrated photodiode mounted on the integrating sphere.}
  \label{flatfield_cryo}
\end{minipage}%
\begin{minipage}{.47\textwidth}
  \centering
  \vspace{-12mm}
  \includegraphics[width=0.89\linewidth]{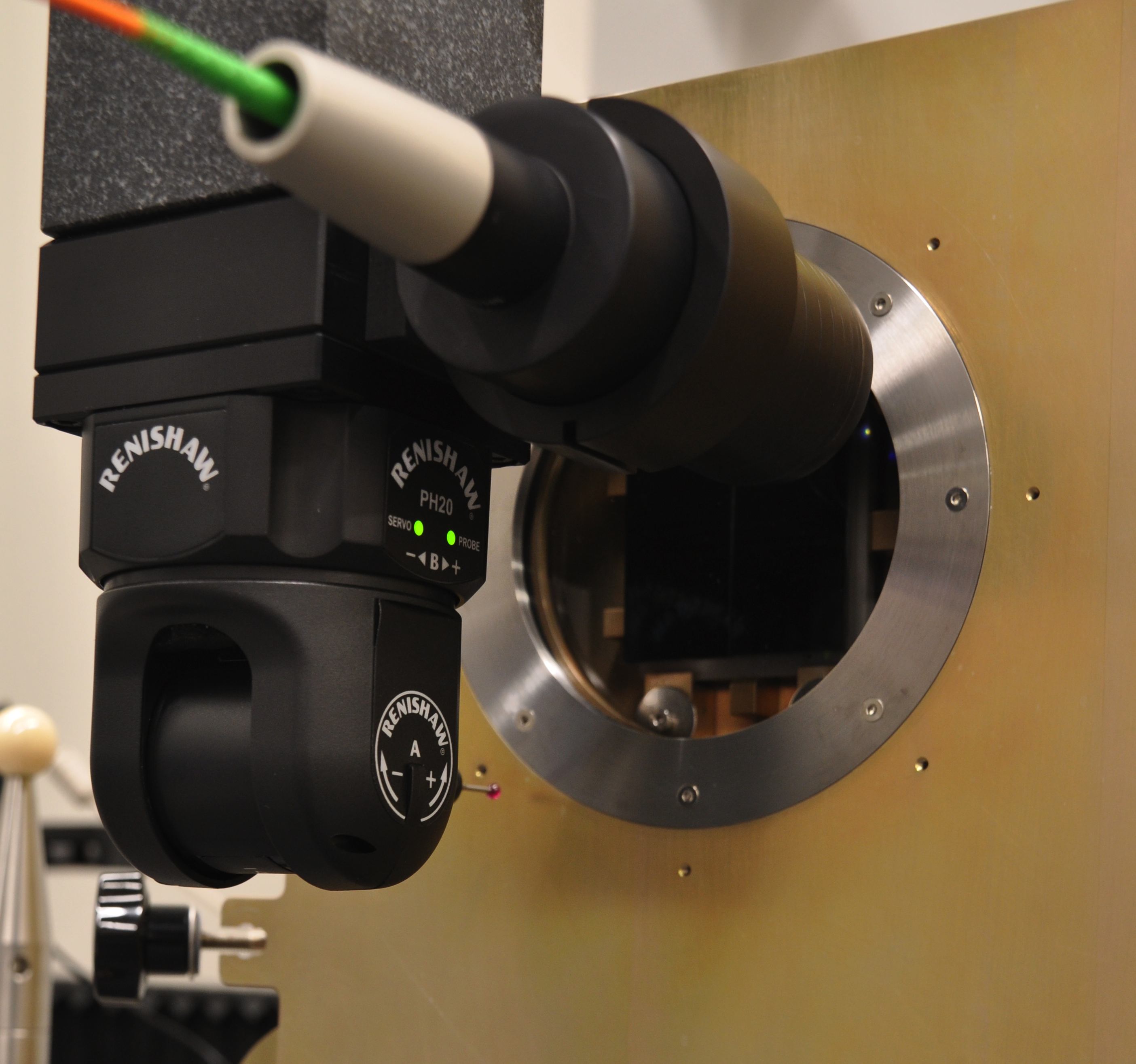}
   \vspace{2mm}
  \captionsetup{width=0.9\linewidth}
  \captionof{figure}{Photo showing the coplanar test setup, with the test cryostat, Micro-Epsilon confocalDT 2405 confocal microscope, mounted in a Wenzel XO-Orbit 87 CMM.}
  \label{coplanar_cryo}
\end{minipage}
\end{figure}

There are two processes in which we need to work with the detectors cold and consequently under a vacuum, in all detector illumination and operation processes, and to also measure the detector topography and coplanarity at operating temperatures. Individual setups are being constructed to perform these two separate functions, a flat fielding setup shown in Fig. \ref{flatfield_cryo}, and a coplanarity setup shown in Fig. \ref{coplanar_cryo}.

%% tcryo configuration photographs

%% flat fielding setup
\subsection{Flat Fielding}	
The flat fielding setup will use a Labsphere\footnote{http://www.labspherestore.com} 300 mm integrating sphere to illuminate detectors mounted in the test cryostat shown in Fig. \ref{flatfield_cryo}. An OSI Optoelectronics\footnote{http://www.osioptoelectronics.com} 44 mm$^{2}$ calibrated photodiode will be mounted on the integrating sphere for flux measurements. The photodiode should produce approximately 2.5 pico-Amps for a flux level of 100 photons per second per 15 $\mu m^{2}$ at 9700 $\AA$. We are using a Keithley\footnote{http://www.keithley.com} 6514 electrometer with triaxial cabling to measure the current delivered from the photodiode. For a light source a Horiba\footnote[1]{http://www.horiba.com/us/en/} iHR 320 monochromator will feed a Fiberguide\footnote[2]{http://www.fiberguide.com} slit to round fiber optic cable, and the round end of the fiber will feed the integrating sphere. In Fig. \ref{clean_hood} a photodiode is mounted in the test dewar at the plane where the detectors will be mounted to calibrate the flux at the detector versus the flux at the calibrated photodiode mounted on the integrating sphere.

%% coplanar setup
\subsection{Coplanar Measurement}
The coplanar test cryostat is mounted to the surface plate of the Wenzel CMM as shown in Fig.\ref{subpixel_positioning}. With the confocal microscope mounted on the CMM's head, as shown in Fig.\ref{coplanar_cryo}, detector surface measurements will be performed at ambient and operating temperatures. With the sub-pixel illuminator mounted on the CMM's head detector characterization will be performed, as can be seen in Fig.\ref{subpixel_positioning}.

%% cold shutter for NIR darks
\subsection{Cold Shutter}	
In the NIR the thermal blackbody flux is quite high, thus to be able to perform dark exposures, with zero incident flux, it is necessary to have a light-tight shutter which operates at or near the detectors operating temperature. Initially the cryostats were designed to use a Phytron\footnote[3]{http://www.phytron-elektronik.de} VSS 32 extreme condition stepper motor which would operate at cryogenic temperatures, but later it was decided to use an external stepper motor with a FerroTec\footnote[4]{http://www.ferrotec.com} ferro-fluidic feed thru. This approach greatly simplified matters as it moved the limit switches and all other electronics outside of the test cryostat allowing for simpler control, and it also gives an external mechanical reference of shutter position.

%% Iron 55 X-ray source				
\subsection{Fe55 Source}
An Fe55 source is useful in detector characterization because it gives a calibrated amount of illumination in approximately a point source. Janesick\cite{Janesick:1985tu} gives the ideal quantum yield for a silicon CCD as \be \eta_{i}=\frac{E_{\lambda}}{3.65}  \/   (\lambda \/ <  \/ 1000\AA) \ee telling us that the fe55 k$\alpha$ X-ray with energy 5.9 keV should yield approximately 1620$e^{-}$ in a single photon event. An Eckert \& Ziegler\footnote[5]{http://www.ezag.com/home/products/isotope-products.html} 10 mCi Fe55 source in an a3204 capsule will be manipulated using a rotary vacuum feed thru. The rotary feed thru will use an arm, adjustable in length, to give access to the entire detector active surface.

%% cryo-cooler
\subsection{Cryo-cooler}	
The PFS cameras will be using the free piston sterling cycle engine CryoTel GT cryocoolers for which an active anti-vibration dampener is currently being developed and tested. As we have a number of detectors to characterize the test cryostat will allow us to run the cooler and anti-vibration motors for extended periods of time.  But also the CryoTel GT with it's ability to easily adjust temperature will ultimately give extra flexibility in finding an optimal operating temperature for our detectors.

%% vacuum system			
\subsection{Vacuum System}
For vacuum generation we will use an Edwards\footnote[6]{http://www.edwardsvacuum.com} turbo molecular pump cart. The pump cart is supplied with a valve that prevents the roughing pump from back streaming into the turbo pump in case of a power loss, but we also desire to isolate the turbo pump from our detectors in the event of a power loss. For this safety valve we will use a pneumatically opened / spring close right angle valve which will be operated  using a 110V electro-pneumatic valve that will turn off the air supply consequently closing the valve upon a power interruption.

%% cleanliness handling procedures
\subsection{Handling Procedures}			
While the detectors must be kept clean, it was decided to perform this work outside of a clean room. The clean room added considerable amounts of overhead to any work due to having to gown and un-gown to perform any task, and it was determined we could maintain a sufficient level of cleanliness without this inconvenience. We will instead use a remote room with low traffic where all of the air vents have been fitted with HEPA filters. We will use tacky mats to keep tracked in dust to a minimum, and wear lab coats, nitrile gloves, bouffants, shoe covers, and face masks while performing all work with the detectors. 

\begin{figure}[h]
         \centering
	\includegraphics[width=0.5\textwidth]{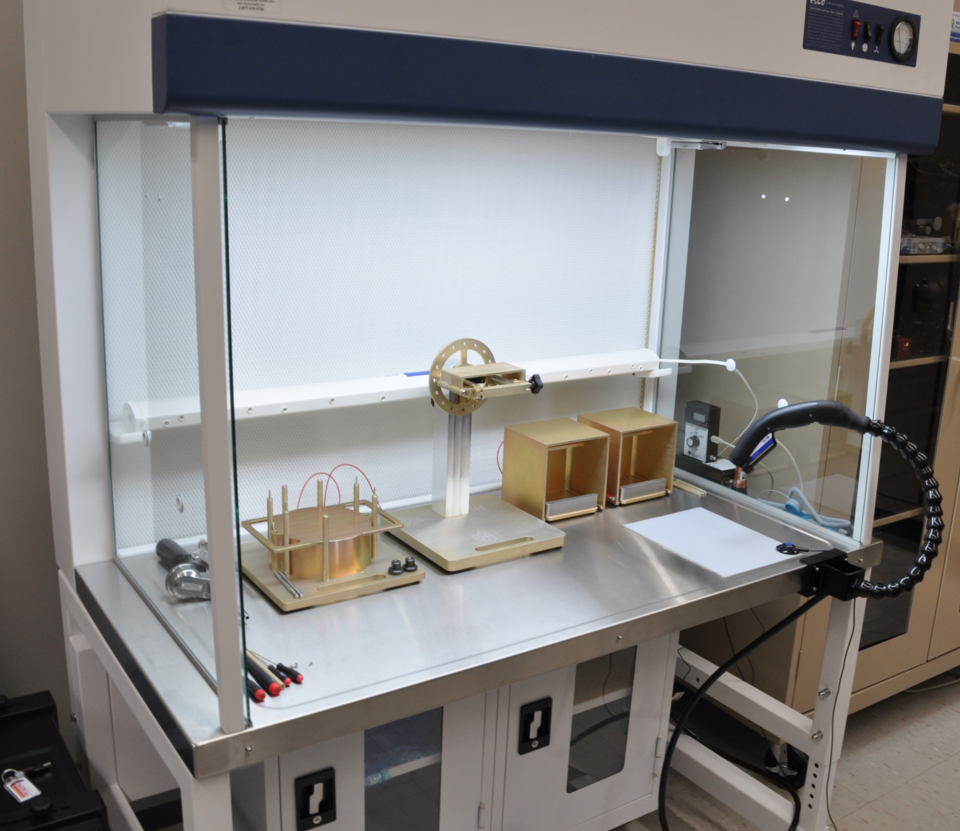}
	\caption{Laminar clean hood with fixtures for assembling detector packages. }
	\label{esco_hood}
\end{figure}

\begin{figure}[h]
         \centering
	\includegraphics[width=0.4\textwidth]{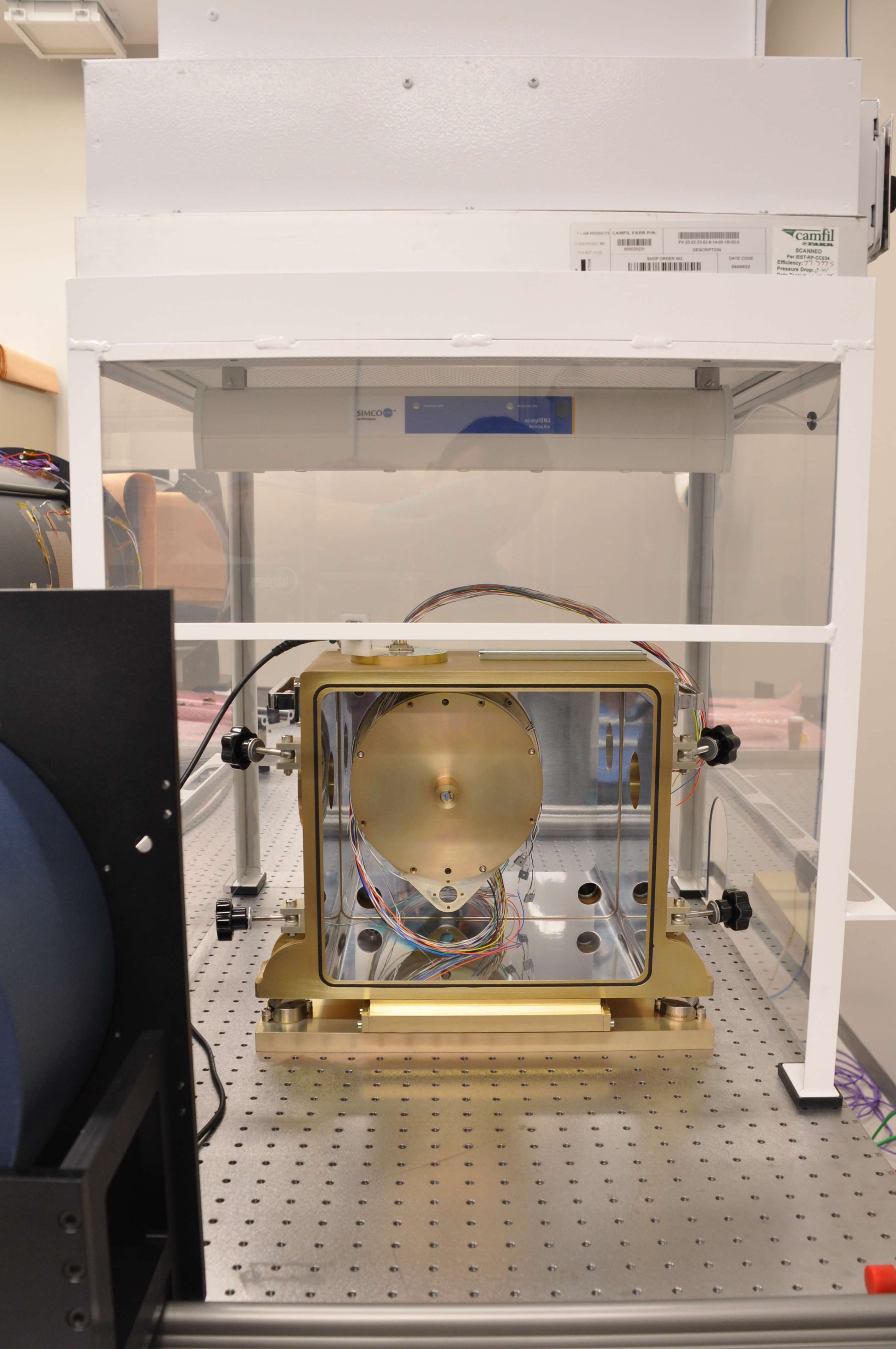}
	\caption{Front view showing the test cryostat open and underneath the portable clean hood. The OSI Optoelectronic photodiode is mounted inside the dewar at the plane where the detectors to be tested will be mounted. }
	\label{clean_hood}
\end{figure}

All open work on the detectors will be performed in a Esco\footnote[7]{http://us.escoglobal.com/laminar-flow-hoods.php} horizontal laminar flow hood shown in Fig.\ref{esco_hood}. The Laminar hood will be used to mount the detector in protective receivers using fixtures specially designed to minimize handling of the detectors and to minimize the number of tools required to perform the task. The receiver, with the detector inside, will be loaded into the vacuum chamber using a portable clean hood. The portable clean hood was constructed using a Clean Air Products\footnote[8]{http://www.cleanairproducts.com} ULPA laminar flow fan filter unit and is shown in Fig. \ref{clean_hood}. 

All electronic devices are sensitive to electro-static damage (ESD), but the Hamamatsu devices have proven to be extremely sensitive. Hardy\cite{Hardy:2012cz} identifies the main risk for damage to the CCDs, it is the connection of the ribbon cable. The ribbon cable has a kapton insulator which holds charge. Since voltages above 15 volts could damage the detectors we desire to have zero surface charge voltage when connecting the ribbon cable.

%% laminar flow hood with handling fixtures

%% dog house clean hood on flat field tcryo

 %%%%%%%%%%%         Sub-Pixel Illumination                   %%%%%%%%%%%%%%%%%%%
\section{Sub-pixel Illumination}\label{sub-pixel}
\begin{figure}[h]
	\centering
	\includegraphics[width=0.9\textwidth]{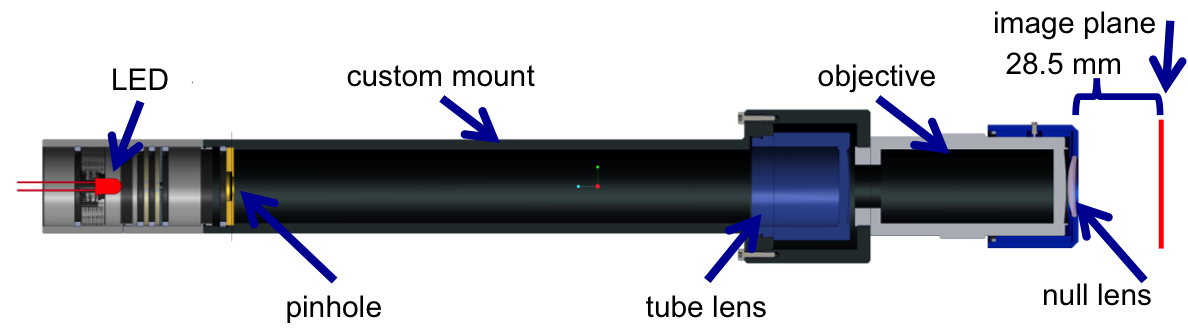}
	\caption{Cross section view of the sub pixel illuminator showing the Mitutoyo NIR objective and tube lens along with pinhole, LED source, and null lens in a custom mount.}
	\label{sub_pixel_xsec_view}
\end{figure}

Having the ability to illuminate a single 15 $\mu$m pixel will allow another mechanism by which to characterize charge diffusion, charge transfer efficiency, and persistence/reciprocity. The original plan was to image a 20 $\mu$m in diameter pinhole with a 10x Mitutoyo\footnote[9]{http://www.mitutoyo.com} infinity corrected objective and tube lens as was used in Fairfield 2006\cite{Fairfield:bb}. The objective and tube lens will be mounted in a custom optical mount shown in cross section in Fig.\ref{sub_pixel_xsec_view}. But we will be imaging this spot through a 13mm thick vacuum window which will introduce spherical aberrations consequently degrading the imaging ability of this system.

%% cross section view of sub pixel illuminator

%% spot characterization
\subsection{Spot Characterization}
\begin{figure}[h]
\centering
\begin{minipage}{.47\textwidth}
  \centering
  \includegraphics[width=\linewidth]{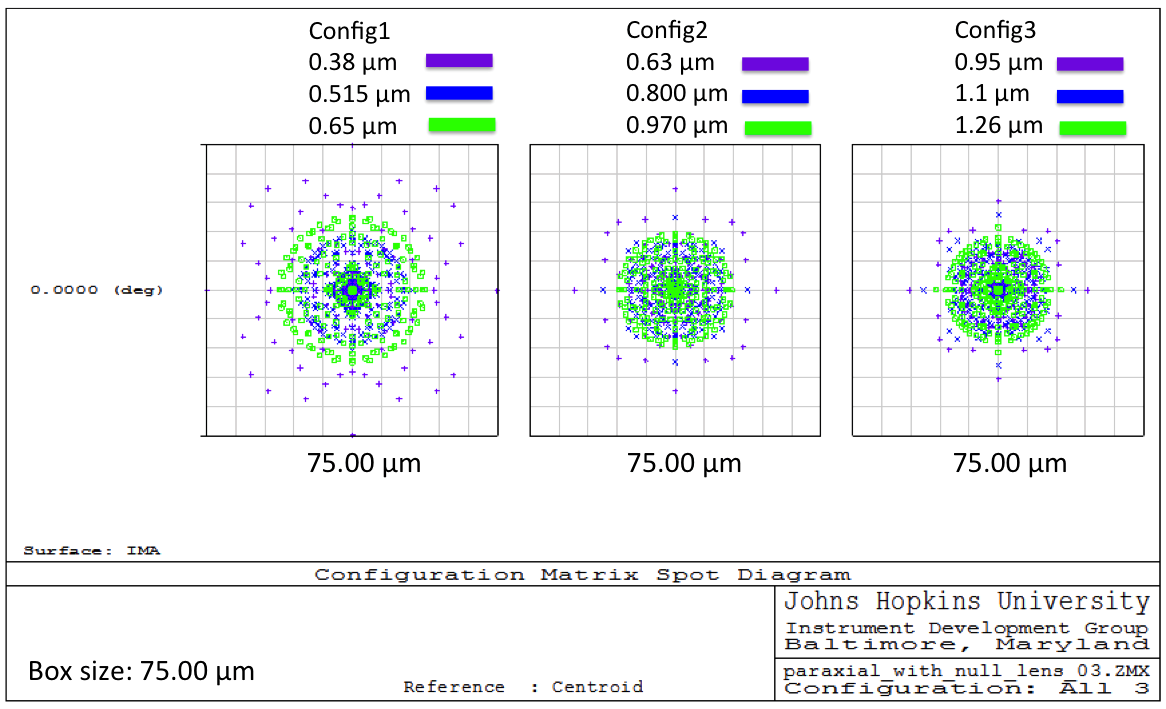}
  \vspace{2mm}
  \captionsetup{width=0.95\linewidth}
  \captionof{figure}{Zemax spot diagram for imaging system without null lens showing that the 13mm vacuum window degrades the spot to nearly 75 $\mu$m in diameter, or nearly five pixels.}
  \label{spot_diagram_wo_null}
\end{minipage}%
\begin{minipage}{.47\textwidth}
  \centering
  \includegraphics[width=\linewidth]{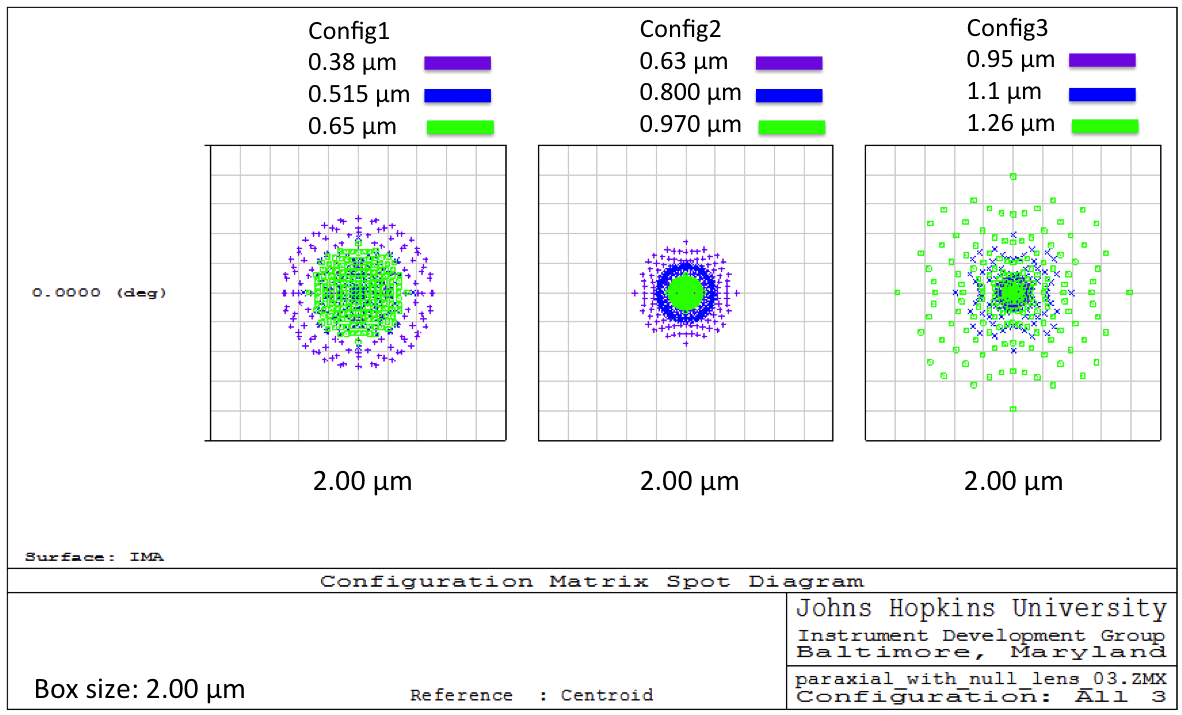}
   \vspace{2mm}
  \captionsetup{width=0.95\linewidth}
  \captionof{figure}{Zemax spot diagram for imaging system with null lens corrector for vacuum window shows that a 2 $\mu$m diameter spot size is optically achievable.}
  \label{spot_diagram}
\end{minipage}
\end{figure}

First Zemax\footnote[1]{http://radiantzemax.com} was used to analytically characterize the imaging ability of an ideal optical system with equivalent numerical aperture and image plane distance. Fig. \ref{spot_diagram_wo_null} is a Zemax spot diagram showing that the addition of the window in the optical path results in a spot size on the order of 75 $\mu$m. Next Zemax was used to design a null lens to correct for the spherical errors introduced by the vacuum window. Fig. \ref{spot_diagram} is the Zemax spot diagram with the addition of the null lens, we can now see that imaging a 2 $\mu$m diameter spot size is acheivable. 

\begin{figure}[h]
\centering
\begin{minipage}{.5\textwidth}
  \centering
  \includegraphics[width=\linewidth]{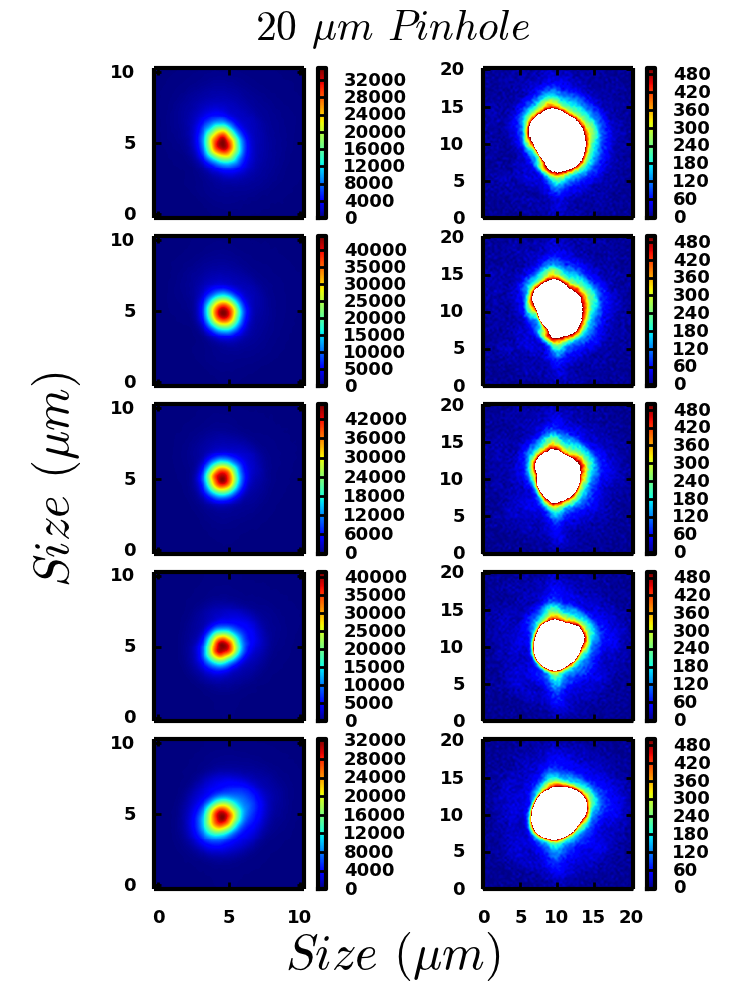}

  \captionsetup{width=0.95\linewidth}
  \captionof{figure}{20 $\mu$m pinhole through focus images, with raw images on the left and thresholded images on the right to highlight the wings of the illumination.}
  \label{20_micron_pinhole_images}
\end{minipage}%
\begin{minipage}{.5\textwidth}
  \centering
  \includegraphics[width=\linewidth]{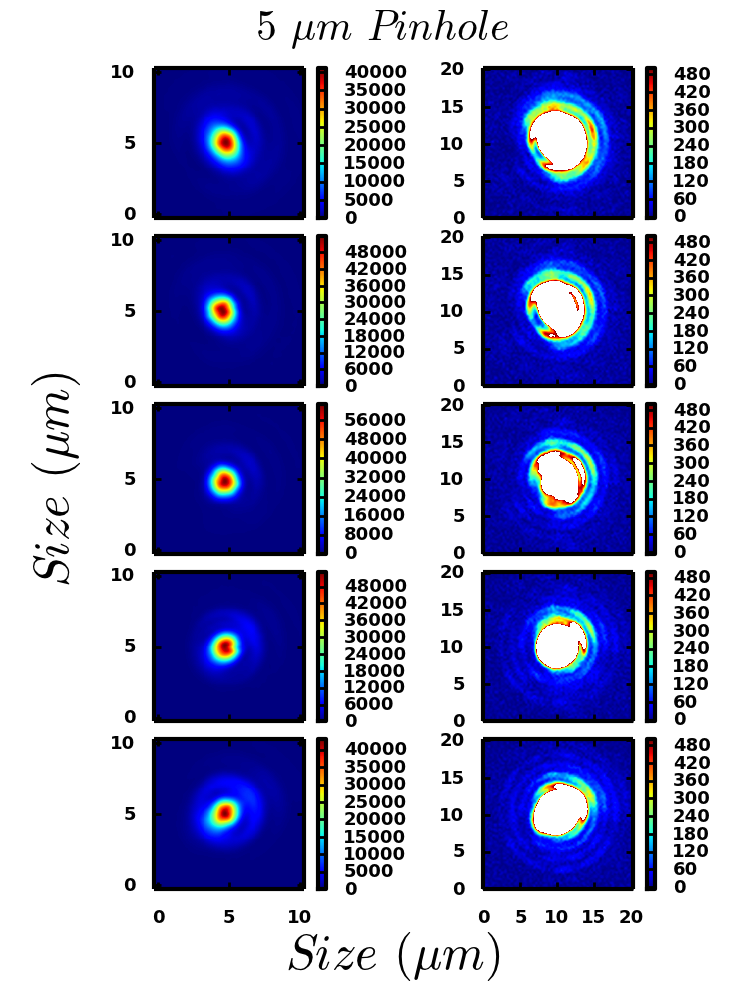}

  \captionsetup{width=0.95\linewidth}
  \captionof{figure}{5 $\mu$m pinhole through focus images, with raw images on the left and thresholded images on the right to highlight the wings of the illumination.}
  \label{5_micron_pinhole_images}
\end{minipage}
\end{figure}

To physically characterize the spot imaged by the sub-pixel illuminator a Melles Griliot\footnote[2]{http://www.mellesgriot.com} 25X objective was used for imaging onto thee detector of a QSI\footnote[3]{http://qsimaging.com} RS3.2 camera with 6.8 $\mu$m pixels. Fig.\ref{20_micron_pinhole_images} is the image of a 20 $\mu$m pinhole taken through focus, and Fig.\ref{5_micron_pinhole_images} is the same set of images taken through focus but now using a 5 $\mu$m pinhole. It can be seen in lowest right hand panel of Fig.\ref{5_micron_pinhole_images} that the imaged spot is an Airy function, while in Fig.\ref{20_micron_pinhole_images} the image is not an Airy function. This implies that the 20 $\mu$m pinhole is not a point source. Also the flux with a 20 $\mu$m pinhole is to great for the final testing where ideally all the flux will go into a single pixel with a full well depth of 64,000 counts.  

%% Zemax spot diagrams

%% spot through focus images

\begin{figure}[h!]
	\centering
	\includegraphics[width=0.6\textwidth]{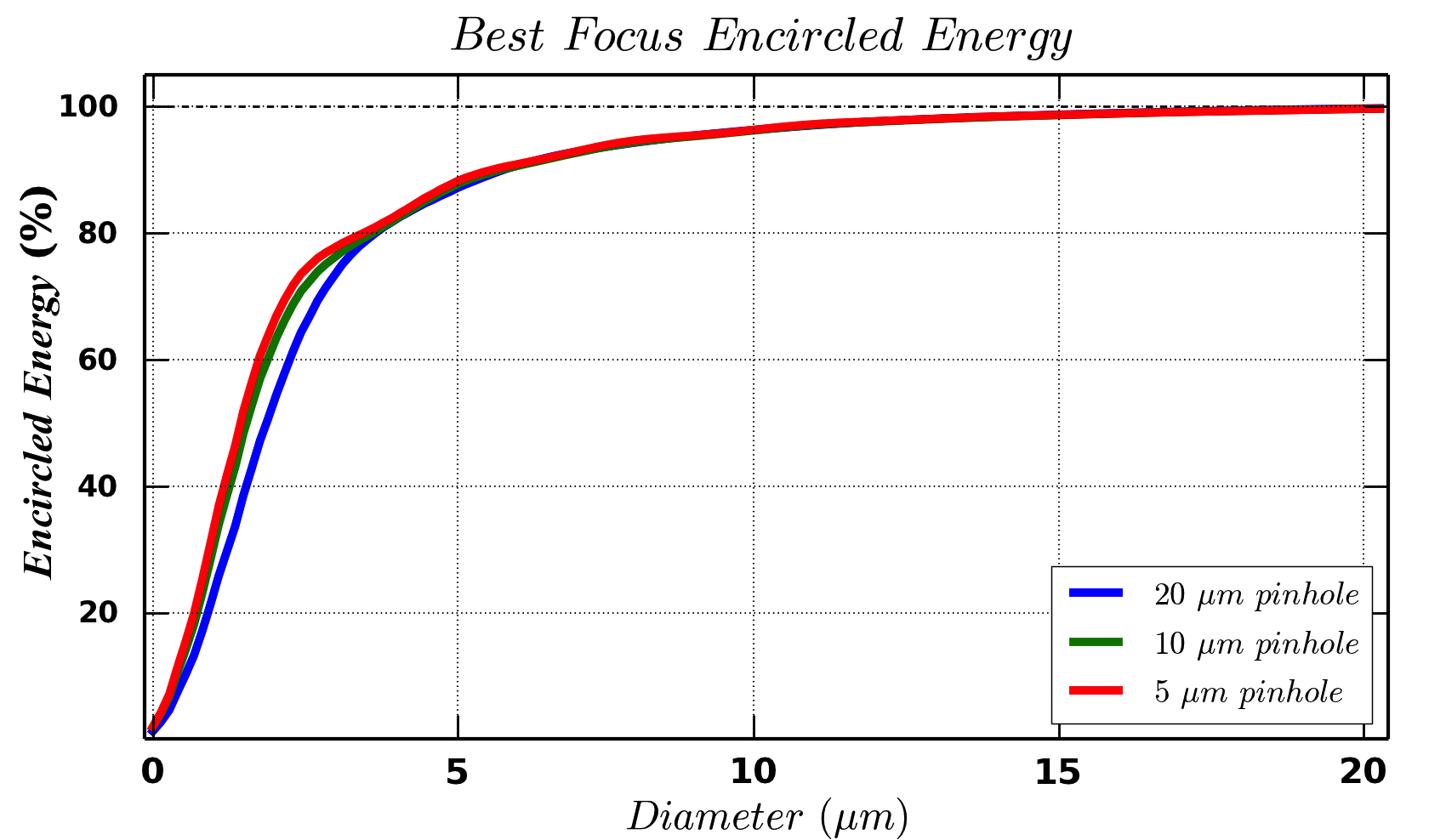}
	\vspace{3mm}
	\caption{Encircled energy versus diameter for 20, 10 and 5 $\mu$m pinhole sizes without the vacuum window. While the smaller pinholes did produce a smaller Airy disc, there was little improvement at diameters greater than 5 $\mu$m.}
	\label{ee_best_focus}
\end{figure}

The first attempts to quantify the imaged spot size used FWHM as the measuring metric, but it was quickly discarded in favor of using 100\% encircled energy. Fig.\ref{ee_best_focus} is a plot of encircled energy as a function of spot diameter for the best focus images of the 20, 10, and 5 $\mu$m pinholes. While an improvement was realized in the diameter of the Airy disc by using a smaller diameter pinhole, the improvements at large diameters were marginal as can be seen in Fig.\ref{ee_best_focus}.

\begin{figure}[h]
	\centering
	\includegraphics[width=0.6\textwidth]{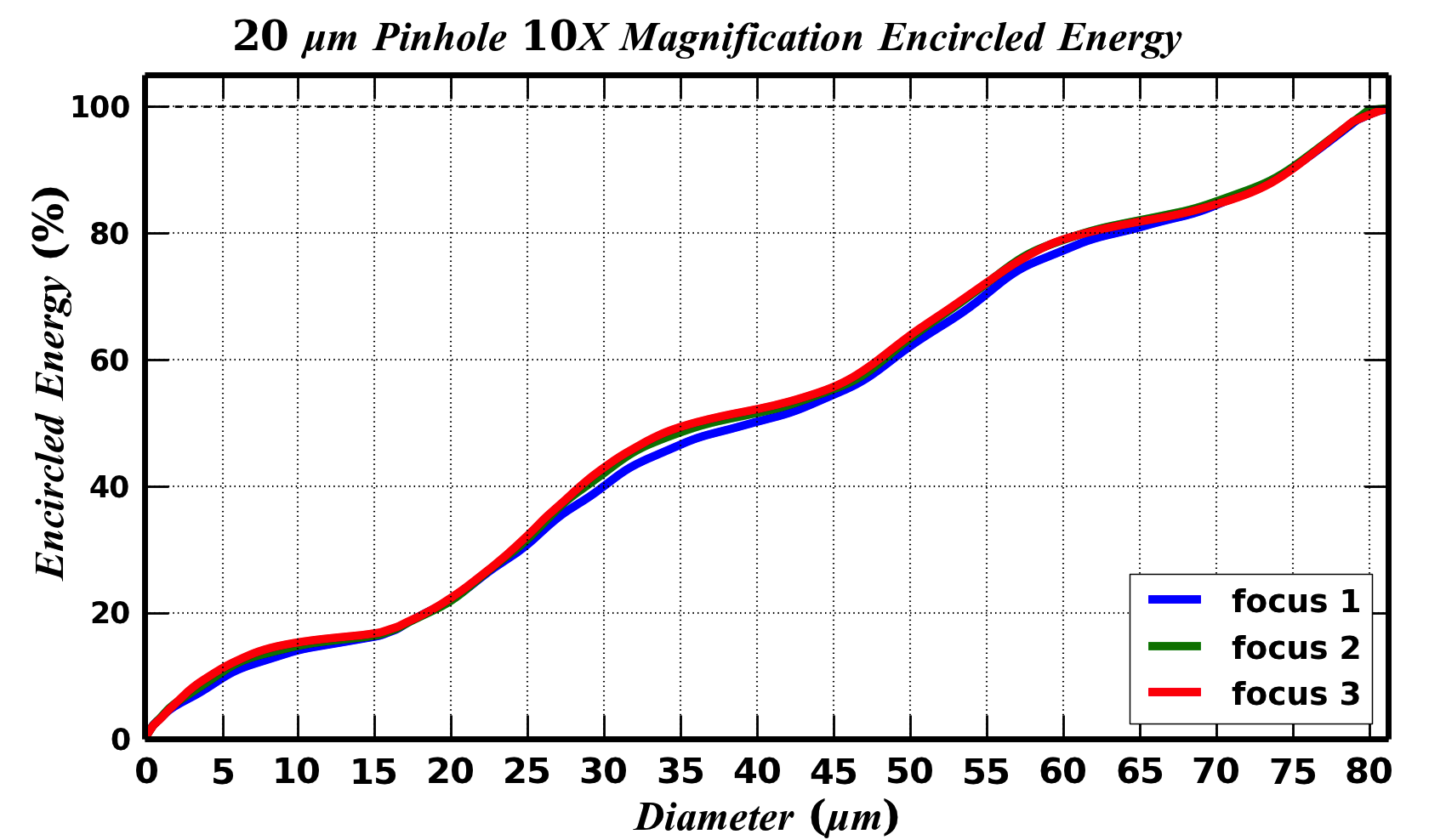}
	\vspace{3mm}
	\caption{The encircled energy diameter through the 13 mm thick vacuum window is increased to approximately 80 $\mu$m, versus 20 $\mu$m without. The optimal focal distance is vaguely defined due to the spherical aberration introduced by the vacuum window.  }
	\label{ee_thru_window_plot}
\end{figure}

Next the encircled energy was measured using a 20 $\mu$m pinhole, but imaging through a 13 mm thick $\sfrac{\lambda}{10}$ vacuum window. Fig.\ref{ee_thru_window_plot} is the encircled energy plot with the addition of the vacuum window. The 100\% encircled energy diameter is now approximately 80 $\mu$m up from 20 $\mu$m without the window. Also the addition of the vacuum window caused the best focus focal distance to be vaguely defined. 

A null lens to correct for the vacuum window has been designed and ordered. The encircled energy will again be measured to quantify the effectiveness of the null lens to correct for the spherical aberrations introduced by the vacuum window. But is has become obvious that imaging a spot smaller than a 15 $\mu$m pixel is not a trivial task. The diameter of the Airy disk is a function of the wavelength and the numerical aperture (NA) of the optics. A smaller spot diameter could be achieved with a higher NA, but this effectively limits the working distance of the optics. 

%% encircled energy best focus without vacuum window

%% encircled energy 20 micron pinhole thru vacuum window

%% spot positioning
\subsection{Spot Positioning}\label{spot_positioning}
\begin{figure}[h]
	\centering
	\includegraphics[width=0.5\textwidth]{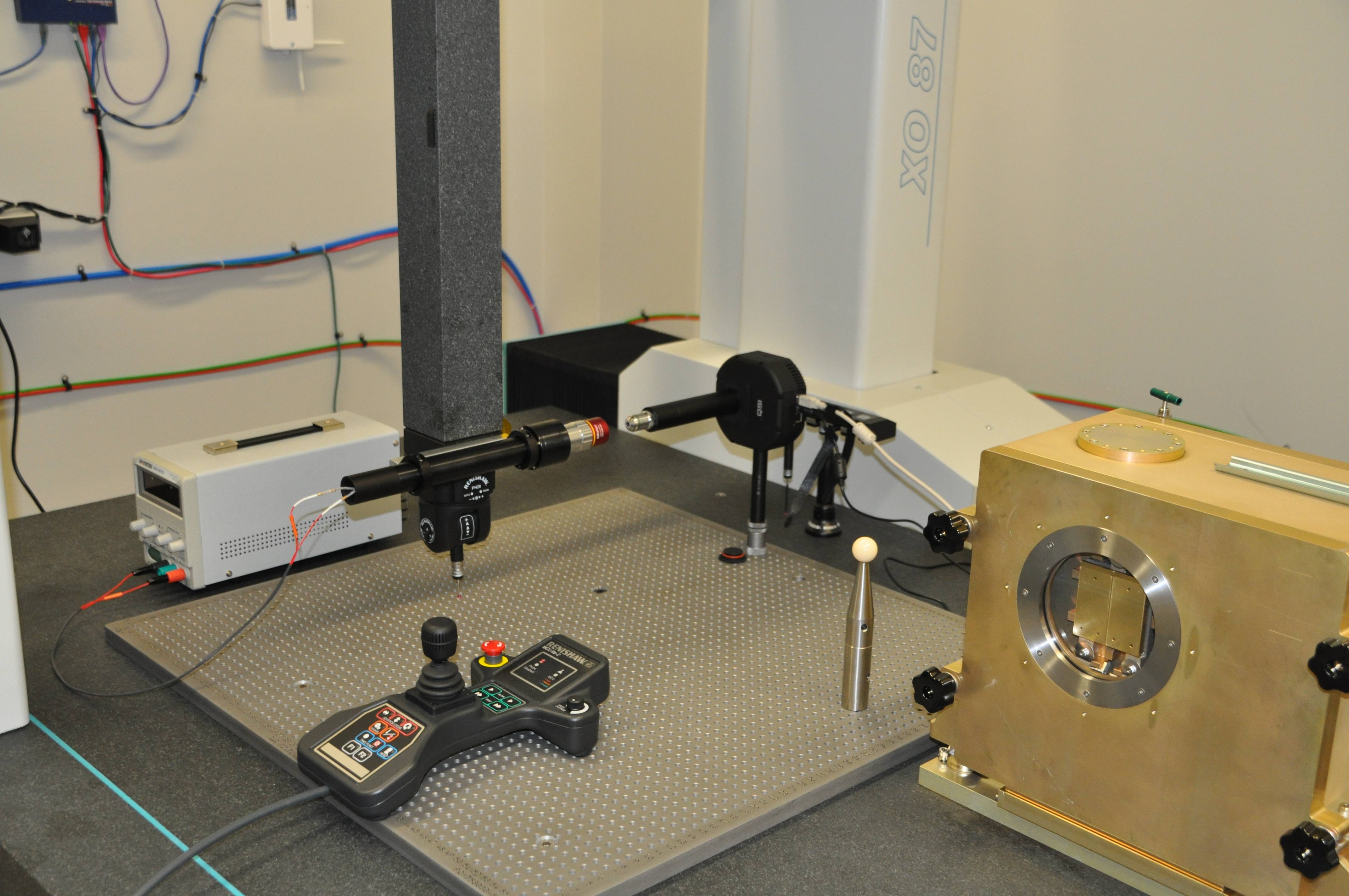}
	\caption{The CMM will be used to position spot on detector inside the dewar. The CMM is also being used to characterize the spot because it allows for the simple positioning of the microscope through a range of foci. }
	\label{subpixel_positioning}
\end{figure}

To use spot illumination effectively we must be able to repeatedly position a 2 $\mu$m diameter spot of light inside a 15 $\mu$m square pixel. The illuminator is mounted on the side of the head of the CMM as shown in Fig.\ref{subpixel_positioning}. Looking at Fig. \ref{positional_error_vs_dwell} the error on absolute positioning on the y-axis of the CMM is on the order of 1 $\mu$m, and the for z-axis it is on the order of 3-5 $\mu$m. While Fig. \ref{mean_dist_vs_dwell} shows that the Y-axis has a small offset in absolute position of a few micron, while the Z-axis has an offset on the order of 15 $\mu$m, but these offsets will be of little consequence. To position the spot in a pixel a series of exposures will be used in which the spot is moved in a single direction in a small step size across a pixel. The exposure with the largest number of counts in the pixel of interest will be considered centered. The process will be repeated for the second axis to find its center, which will be performed on the center found for the previous axis.

%% spot illuminator mounted on CMM

 %%%%%%%%%%%      CHARACTERIZATION               %%%%%%%%%%%%%%%%%%%
\section{Detector Characterization}
\begin{figure}[htp]
	\centering
	\includegraphics[width=0.8\textwidth]{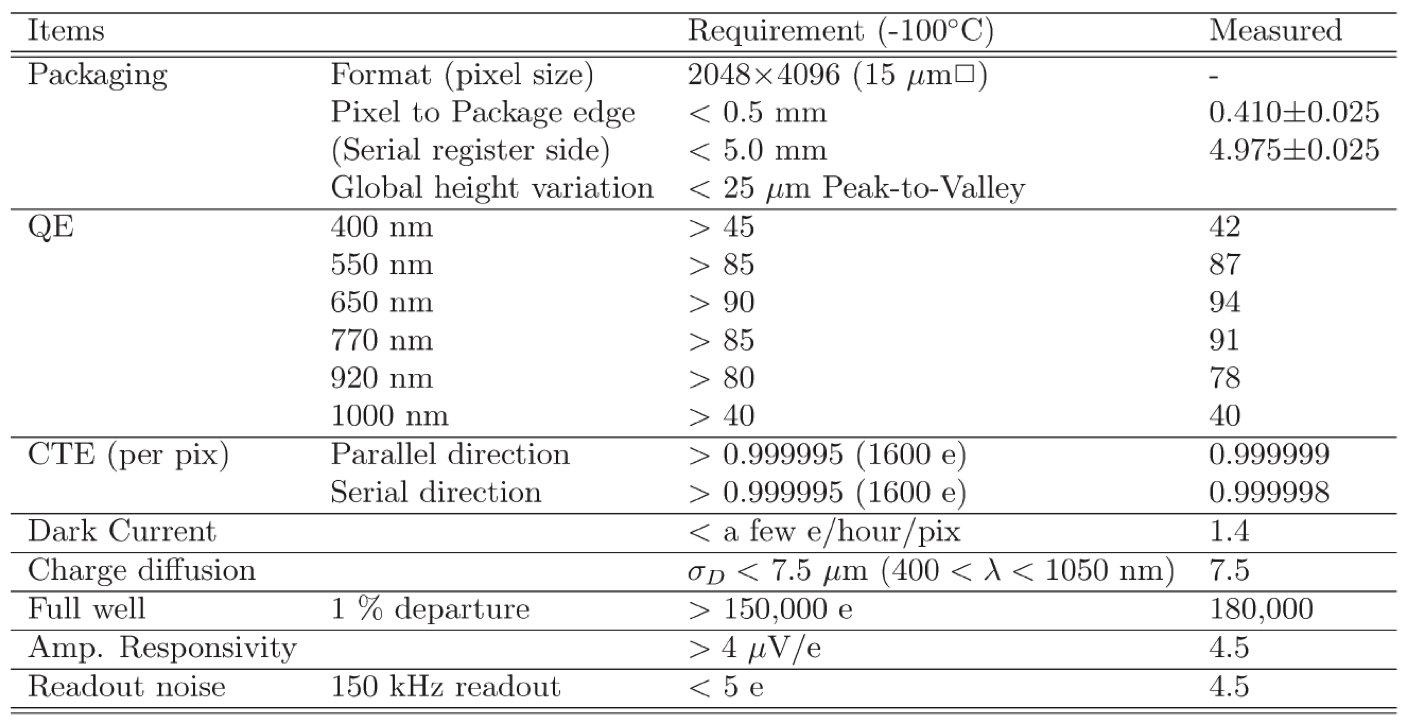}
	\caption{Hyper Suprime-Cam (HSC) detector specifications as given by Kamata\cite{Kamata:2010ic} for the Hamamatsu deep depletion CCDs. }
	\label{detector_specs}
\end{figure}

The HSC CCD specifications are shown in Fig.\ref{detector_specs} as given by Kamata\cite{Kamata:2010ic} are a starting point for PFS. The PFS requirements are much more stringent than for HSC though, as HSC is an imaging instrument which delivers much higher signal to noise ratio to the detector on a per pixel basis. PFS will use a full well depth of 64,000 electrons versus HSC's 150,000 electron full well depth, and PFS will also use a slower 75 kHz readout speed to achieve a read noise of less than 3 electrons.

%% HSC detector specs

%% flat field illumination
\subsection{Photon Transfer Curve (PTC)}	
The photon transfer curve (PTC) is a measure of a detectors response, and is a technique originally pioneered by Janesick\cite{Janesick:1985tu}. The PTC is characterized by Eqn.\ref{eq:PTC} which shows the detectors variance is a linear function in signal. \be \label{eq:PTC}  \sigma^{2}_{s}=\frac{1}{K}S \ee The PTC can be used to measure read noise, dark current, quantum yield, full well, linearity, pixel non-uniformity, sensitivity, signal to noise ratio, offset, and the dynamic range of a detector. A total noise photon transfer curve is a plot of log noise as a function of log signal, and this plot is characterized by three noise regions. The first noise region at zero illumination represents the combined read noise of the detector and electronics, and this region has a slope of zero. The second region is the shot noise dominated regime with a slope of $\sfrac{1}{2}$, and represents the uncertainty in photon counting statistics which go as the square root of the signal. The final region is fixed pattern noise dominated. Fixed pattern noise represents pixel non-uniformity in response and scales proportional to the signal with a slope of 1. 

In the calculation of the variance using Eqn.\ref{eq:VAR}, we are able to remove pixel-to-pixel non-uniformity by simply taking difference between flat field images on a per pixel basis. Also note that Eqn.\ref{eq:VAR} implies that the noise scales as $\sfrac{1}{\sqrt{N_p}}$, and by binning pixels a sufficient level of signal to noise will still be achieved while using shorter exposure times.

\be \label{eq:VAR}  VAR = \sigma^{2} = \frac{\sum_{i=0}^{N_p} \/ ([X_i - \bar X]-[Y_i - \bar Y])^2}{2 \times N_p} \ee

%% QE - quantum efficeiny
\subsection{Quantum Efficiency (QE)}
The QE of the Hamamatsu and Teledyne devices are tested before delivery. We desire to verify the manufactures QE measurements, and we will perform our own QE measurements using flat field exposures. The OSI Optoelectronics photodiode mounted on the integrating sphere will be our test reference QE standard. The uncertainty in the QE calibration of the photodiode is less than $\pm$3\% for wavelengths between 2500 $\AA$ and 10500 $\AA$, and $\pm$5.5\% for wavelengths between 10500 $\AA$ and 11000 $\AA$.

%% dark current
\subsection{Dark Current}
For low illumination applications it is necessary to minimize dark current to achieve good signal to noise, which is accomplished by using lower detector operating temperatures. But the optimal operating temperature is  a tradeoff between minimizing dark current, minimizing persistence decay times, freezing out hot pixels, and maximizing QE. To minimize the effects from dark current and hot pixels a colder operating temperature is desired. While the characteristic decay time for persistence decreases with increasing temperature, and the QE is improved by a few percent in the NIR by operating at warmer temperatures. But for PFS the primary consideration will be to minimize the dark current to maximize the signal to noise ratio of the instrument.  The Cryotel-GT cryocooler will allow for simple adjustment of operating temperature, which will aid in the optimization of the final operating temperature of the detectors.

%% Iron 55  Charge diffusion testing
\subsection{Charge Diffusion}
X-ray photons created by an Fe55 source are much higher energy than the longer wavelength visible to NIR photons that the detectors are designed to collect. One key difference in using a higher energy photon for testing is the penetration depth which scales inversely with the frequency of the photon, i.e. a higher energy photon will deposit itself at a shallower depth than a lower energy longer wavelength photon. Thus to measure charge diffusion it is necessary to generate a large number of events of which a few will be at a sufficient enough penetration depth to be comparable to a lower energy photon. Rodney\cite{Rodney:2006dr} gives a method by which to analyze a large number of Fe55 events to measure the charge diffusion properties of a detector. Another technique we will use to measure charge diffusion is the sub-pixel illuminator. We will sample a handful of pixels through a range of illumination levels, and convolve the images with gaussians to measure the charge spread $\sigma$.

%% Iron 55 CTE 
\subsection{Charge Transfer Inefficiency (CTI)}
Since the Fe55 X-ray delivers a quantified dosage it is extremely useful for measuring charge transfer inefficiency. X-ray photon events will be collected, and the number of electrons arriving to the amplifiers will be measured as a function of the number of transfers. Flat fields can also be used to measure CTE, and Marshall\cite{Marshall:2001vt} outlines a technique extended pixel edge response (EPER) in which the detector is overclocked to measure the deferred charge.

%% Tests to be performed with sub-pixel illumination charge diffusion and persistence			
\subsection{Non-Linear Photon Transfer Curve}			
We also desire to better understand the correlation between pixels and the non-linear PTC in both the Hamamatsu CCDs and the Teledyne CMOS devices. Downing\cite{Downing:2006wt} shows that the non-linearity is due to spatial correlation between pixels, with a higher correlation down a row rather than across the columns. This effect is on the order of a few percent between pixels, but is on the order of 10\% for the summed correlation of a single pixel. This consequently leads to a 10\% overestimation of the gain. To more accurately measure the gain we will use the autocorrelation variance outlined by Downing\cite{Downing:2006wt}.

%% persistence
\subsection{Persistence-Reciprocity}
Smith\cite{Smith:2008by} gives a model which describes the behavior of persistence in HgCdTe detectors. Teledyne claims to have mitigated persistence in their latest HgCdTe devices. But of a larger concern for PFS is reciprocity, the reduction of quantum efficiency at low illumination levels due to persistence. To quantify the reciprocity we will take one long exposure, and stack many short exposures which sum to the same integration time of the long exposure, and then take the difference of the two images.

 %%%%%%%%%%%     Summary              %%%%%%%%%%%%%%%%%%%
\section{Summary}
We have described here the preparations to enable the metrology and characterization on the detectors for PFS. Multiple testing setups are being constructed to measure focal plane array flatness and to characterize the performance of the detectors. PFS will be using both CCDs and CMOS devices which has required that the testing setups be flexible in their application.

 %%%%%%%%%%%    Acknowledgements              %%%%%%%%%%%%%%%%%%%
\acknowledgments
We gratefully acknowledge support from the Funding Program for World-Leading Innovative R\&D on Science and Technology (FIRST) ''Subaru Measurements of Images and Redshifts (SuMIRe)'', CSTP, Japan.

 %%%%%%%%%%%       BIBLIOGRAPHY                 %%%%%%%%%%%%%%%%%%%
\bibliography{spie_2014_biblio_1}   
\bibliographystyle{spiebib}

 %%%%%%%%%%%      THE END             %%%%%%%%%%%%%%%%%%%
\end{document}